\documentclass[letterpaper,english,footbib]{revtex4}
\usepackage[T1]{fontenc}
\usepackage[latin9]{inputenc}
\usepackage{xcolor}
\usepackage{array}
\usepackage{amsmath}
\usepackage{graphicx}
\PassOptionsToPackage{normalem}{ulem}
\usepackage{ulem}

\makeatletter

\providecommand{\tabularnewline}{\\}

\providecolor{lyxadded}{rgb}{0,0,1}
\providecolor{lyxdeleted}{rgb}{1,0,0}

\@ifundefined{textcolor}{}
{%
 \definecolor{BLACK}{gray}{0}
 \definecolor{WHITE}{gray}{1}
 \definecolor{RED}{rgb}{1,0,0}
 \definecolor{GREEN}{rgb}{0,1,0}
 \definecolor{BLUE}{rgb}{0,0,1}
 \definecolor{CYAN}{cmyk}{1,0,0,0}
 \definecolor{MAGENTA}{cmyk}{0,1,0,0}
 \definecolor{YELLOW}{cmyk}{0,0,1,0}
 }

\makeatother
\usepackage{babel}
\begin{document}
\title{Optimal Vertex Cover for the Small-World Hanoi Networks}
\author{Stefan Boettcher}
\homepage{http://www.physics.emory.edu/faculty/boettcher/}
\affiliation{Physics Department, Emory University, Atlanta, GA 30322; USA}
\author{Alexander K. Hartmann}
\homepage{http://www.compphys.uni-oldenburg.de/}
\affiliation{Institut für Physik, Universität Oldenburg, 
D-26111 Oldenburg; Germany}

\date{\today}

\begin{abstract}
The vertex-cover problem on the Hanoi networks HN3 and HN5 is analyzed
with an exact renormalization group and parallel-tempering Monte Carlo
simulations.  The grand canonical partition function of the equivalent
hard-core repulsive lattice-gas problem is recast first as an
Ising-like canonical partition function, which allows for a closed set
of renormalization-group equations. The flow of these equations is
analyzed for the limit of infinite chemical potential, at which the
vertex-cover problem is attained. The relevant fixed point and its
neighborhood are analyzed, and non-trivial results are obtained both,
for the coverage as well as for the ground-state entropy density,
which indicates the complex structure of the solution space. Using
special hierarchy-dependent operators in the renormalization group and
Monte Carlo simulations, structural details of optimal configurations
are revealed. These studies indicate that the optimal coverages (or
packings) are not related by a simple symmetry.  Using a clustering
analysis of the solutions obtained in the Monte Carlo simulations, a
complex solution space structure is revealed for each system
size. Nevertheless, in the thermodynamic limit, the solution landscape
is dominated by one huge set of very similar solutions.
\end{abstract}
\maketitle

\section{Introduction\label{sec:Introduction}}

We study the vertex-cover
problem~\cite{cover2000,phase-transitions2005} on the recently
introduced set of Hanoi networks~\citep{SWPRL,SWN,SWlong}%
\footnote{Unfortunately, we had to learn that there already exists a
  hierarchical graph with that name that is, in fact, similar but
  otherwise unrelated to the networks discussed here, see
  http://mathworld.wolfram.com/HanoiGraph.html.}. An optimal vertex
cover attempts to find the smallest set of vertices in a graph such
that every edge in the graph connects to at least one vertex in that
set. It is one of the classical NP-hard combinatorial optimization
problems discussed in Ref.~\citep{Karp72}.  The problem is equivalent
to a hard-core lattice gas~\citep{Weigt01}, in which any pair of
particles must be separated by at least an empty lattice site.  The
vertex-cover problem has recently attracted much attention in physics,
because in ensembles of Erd\"{o}s-R\'eny random
networks~\cite{erdoes1960}, phase transitions in the structure of the
solution landscape were found that coincide with a
polynomial-exponential change of the running time of exact
algorithms~\citep{cover2000,phase-transitions2005}.

During the past decade, alternative ensembles of random networks have
attracted the attention of physicists. Well-known examples are
Watts-Strogatz small-world networks~\citep{Watts98} and scale-free
networks~\citep{Barabasi99,Andrade05,Hinczewski06,Zhang07}.  These
networks exhibit more structure and describe the behavior of real
networks much better than Erd\"{o}s-R\'eny
networks~\citep{NewmanSIAM03}.  Also, physical systems (such as the
Ising model~\cite{barrat2001, aleksiejuk2002}) that exist on these
more complex network or lattice structures behave differently compared
to regular (hyper-cubic) lattices or random networks.

Hanoi networks mimic the behavior of small-world systems without the
usual disorder inherent in the construction of such networks. Instead,
they attain these properties in a recursive, hierarchical manner that
lends itself to exact real-space renormalization~\citep{Plischke94}.
These networks do not possess a scale-free degree distribution; they
are, like the original small worlds, of regular degree or have an
exponential degree distribution. These Hanoi networks have a more
physically desirable geometry~\citep{barthelemy_spatial_2010}, with a
mix of small-world links and a nearest-neighbor backbone
characteristic of lattice-based models~\citep{SWN}.

\begin{figure}
\hfil\includegraphics[bb=120bp 230bp 680bp
  660bp,angle=0,clip,scale=0.43]{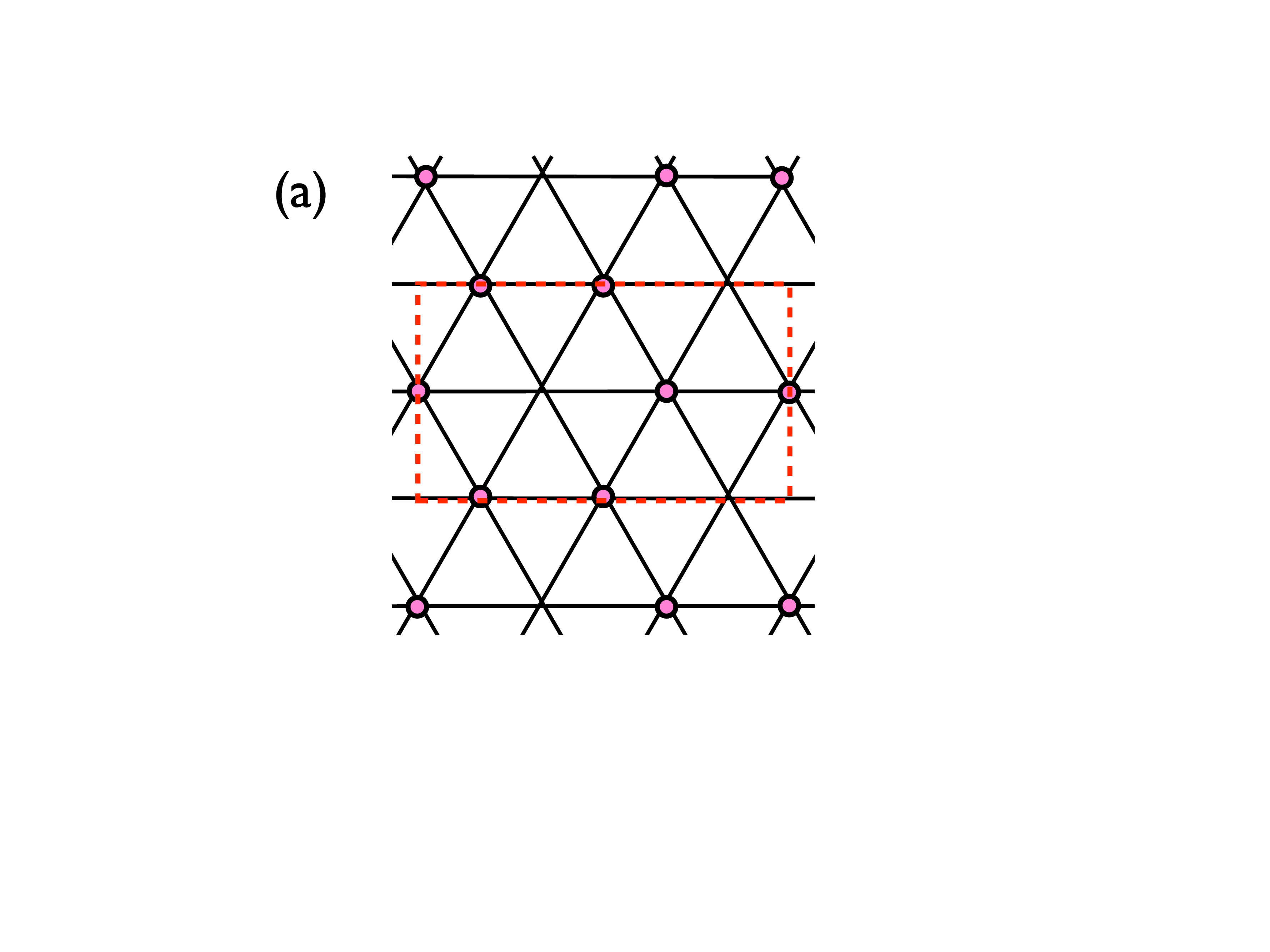}\hfil\includegraphics[bb=120bp
  300bp 660bp 780bp,clip,scale=0.4]{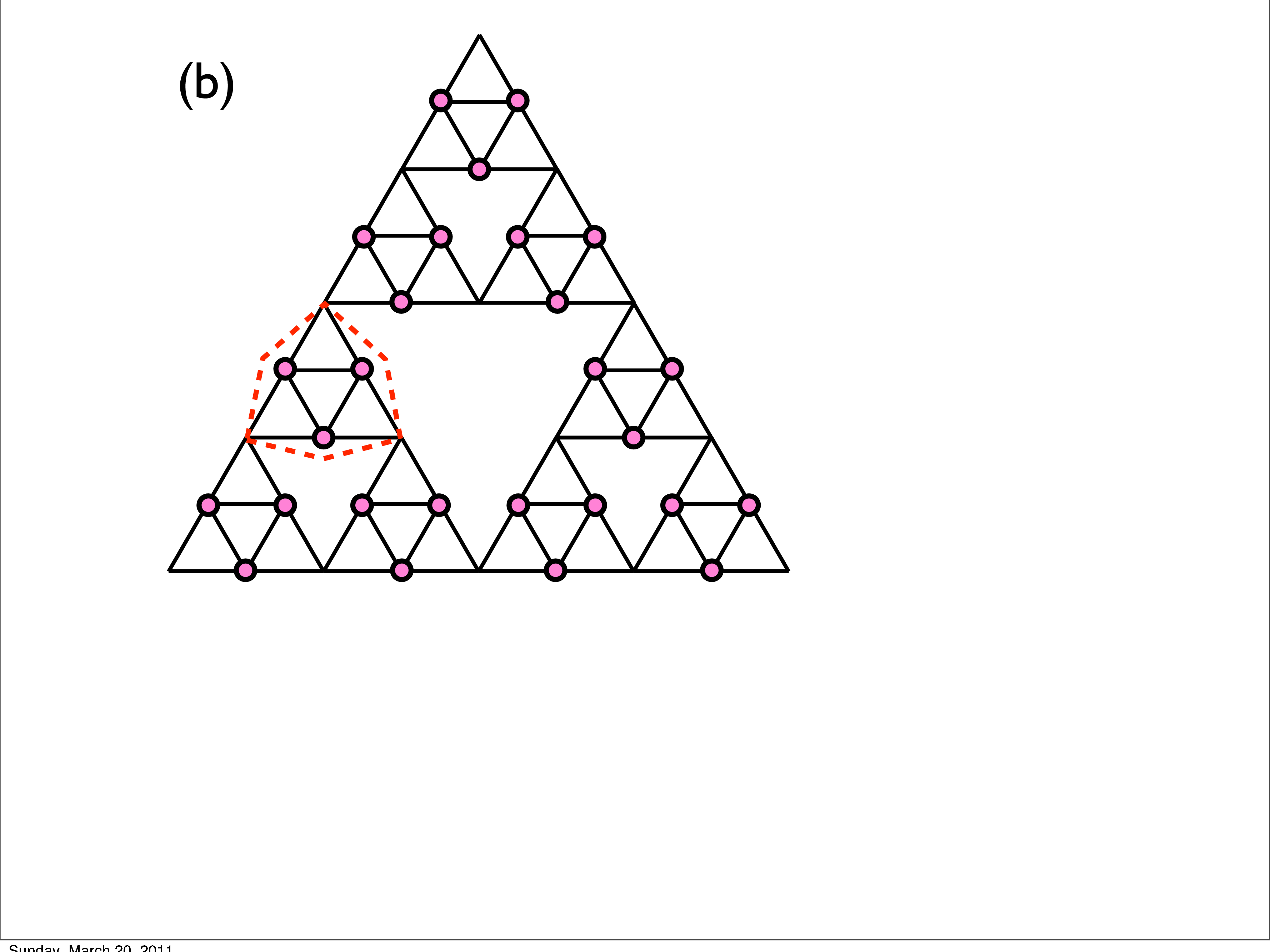}\hfil
\caption{\label{fig:SierpinskiVC}(Color online)
Vertex covering for (a) a triangular lattice and (b) a Sierpinski
gasket. In both cases, the optimal coverage (large dots) is imperfect
(i.e., some edges possess double coverings). Yet, these solutions
either are unique, as for the Sierpinski gasket, or possess a finite
symmetry, such as the possible translations on the triangular lattice,
both cases leading to a vanishing entropy density. For both lattices
it is easily seen that the asymptotic coverage is $\frac{2}{3}$. In
the case of the triangular lattice, the unit cell (dashed red box)
contains two vertices completely and shares half of eight vertices
with other cells, i.e., it has effectively $2+\frac{8}{2}=6$ vertices
of which $1+\frac{6}{2}=4$ are covered. The unit cell in the
Sierpinski gasket contains $3+\frac{3}{2}$ vertices of which the three
fully contained ones must be covered.}
\end{figure}

For the vertex-cover problem considered here, or the equivalent
hard-core lattice gas, it is difficult to find metric structures with
a non-trivial solution. For instance, hyper-cubic lattices are
bipartite graphs that always have an obvious unique and trivial
solution without any conflicts. Of the planar lattices, the triangular
one is certain to exhibit imperfect solutions (i.e., there will be
edges requiring multiple coverings for any solution), but any such
solution is translationally invariant and can be easily enumerated,
leading to a vanishing entropy density. Similarly, a fractal lattice
such as the Sierpinski gasket, say, only has trivial solutions of that
sort. Both of these examples are given in
Fig.~\ref{fig:SierpinskiVC}. In contrast, we find an extensive
ground-state entropy here, similar to the anti-ferromagnet on a
triangular lattice~\citep{Wannier50}. Yet, our ground states do not
appear to be the result of any symmetry relation.  Thus, the study of
the vertex-cover problem on the Hanoi networks affords simple,
analytically tractable examples of coverages that have nontrivial
entropy densities. In fact, analytically we found merely an
approximate algorithm to generate (and enumerate) the set of all
solutions whose true cardinality we can determine at any finite system size
only by exact renormalization.

Using branch-and-bound algorithms, we enumerate exact
solutions~\citep{phase-transitions2005}; however, due to the
exponentially growing running time of this exact algorithm, we are
restricted to rather small system sizes. Hence, for most of the
numerical studies performed here, we use Monte Carlo
simulations~\citep{Newman99b} to generate the solutions and clustering
algorithms to elucidate their correlations~\citep{jain1988}.

Previous work~\citep{Weigt01} has focused on averaged properties on
locally tree-like (mean-field) networks using the replica method,
unearthing interesting phase transitions for the problem. Thus far,
there are only a few investigations into the statistical mechanics of
the vertex-cover problem on more complex networks. In a study of
randomly connected tetrahedra~\citep{Weigt03}, glassy behavior was
observed. When introducing degree-correlations, it was found that the
vertex-cover problems becomes numerically harder~\cite{vazquez2003}.

This paper is organized as follows. In Sec.~\ref{sec:Graph-Structure}
we review the properties of the Hanoi networks. In
Sec.~\ref{sec:Vertex-Cover-Problem} we briefly recount the relevant
theory for a thermodynamic study of vertex cover in terms of a
hard-core lattice gas. In Sec.~\ref{sec:RG-for-the}, we develop the
renormalization group treatment of the lattice gas, with most of the
technical details deferred to the Appendix~\ref{sec:Appendix}, and its
application to the Hanoi networks HN3 and HN5. A detailed numerical
study of the problem follows in
Sec.~\ref{sec:Monte-Carlo-Simulations}.  We present our conclusions
and an outlook for future work in Sec.~\ref{sec:Conclusions}.

\section{Geometry of the Hanoi Networks\label{sec:Graph-Structure}}

\begin{figure}
\includegraphics[bb=45bp 20bp 822bp 480bp,clip,scale=0.3]{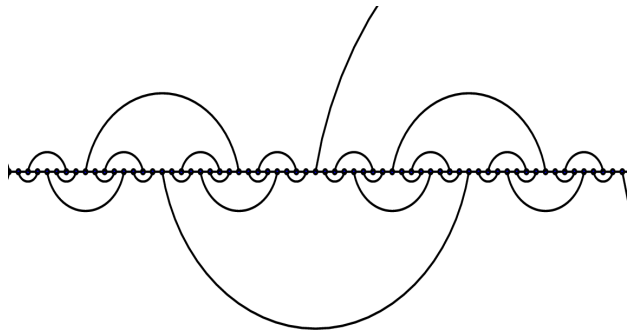}
\caption{\label{fig:3hanoi}
Depiction of the 3-regular network HN3 on a semi-infinite line. Note
that HN3 is planar.}
\end{figure}

Each of the Hanoi networks possesses a simple geometric backbone, a
one-dimensional line of sites $0\leq n<N=2^{k}+1$~\citep{SWPRL,SWN}. Most importantly, all sites are connected to
their nearest neighbors, ensuring the existence of the
$1d$-backbone. To generate the small-world hierarchy in these
networks, consider parameterizing any integer $n$ (except for zero)
uniquely in terms of two other integers $(i,j)$, $i\geq1$, via
\begin{eqnarray}
n & = & 2^{i-1}\left(2j+1\right),
\label{eq:numbering}
\end{eqnarray}
where $i$ denotes the level in the hierarchy and $j$ labels
consecutive sites within each hierarchy. For instance, $i=1$ refers to
all odd integers, $i=2$ to all integers once divisible by 2 (i.e.,
2, 6, 10,...), and so on. In these networks, aside from the backbone,
each site is also connected with some of its neighbors \emph{within}
the hierarchy. For example, we obtain a 3-regular network HN3 (best
done on a semi-infinite line) by connecting first the backbone, then 1
to 3, 5 to 7, 9 to 11, etc, for $i=1$, next 2 to 6, 10 to 14, etc, for
$i=2$, and 4 to 12, 20 to 28, etc, for $i=3$, and so on, as depicted
in Fig.~\ref{fig:3hanoi}. Previously~\citep{SWPRL}, it was found that
the average chemical path between sites on HN3 scales as
\begin{equation}
d^{HN3}\sim\sqrt{l}
\label{eq:3dia}
\end{equation}
with the distance $l$ along the backbone. 

\begin{figure}
\includegraphics[bb=120bp 100bp 400bp 700bp,clip,angle=-90,scale=0.5]{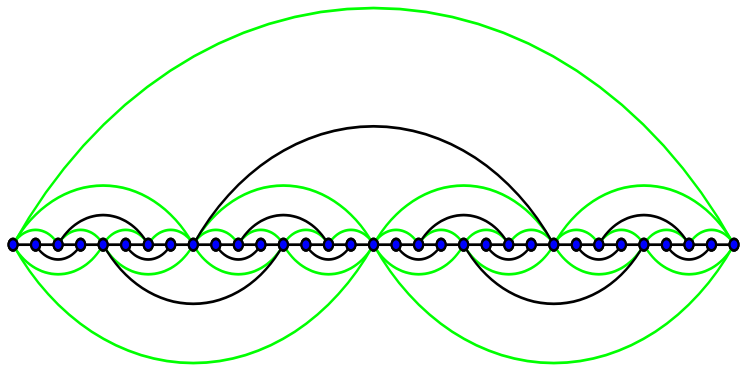}
\caption{\label{fig:5hanoi}(Color online)
Depiction of the planar network HN5, consisting of an HN3 core (black
lines) with the addition of farther-reaching long-range edges (shaded
lines). Note that sites on the lowest level of the hierarchy have
degree 3, then degree 5, 7, etc, comprising a fraction of $1/2,$
$1/4$, $1/8$, etc., of all sites, which makes for an average degree of 5
in this network. (There is no distinction made between black and
shaded lines in our studies here.)}
\end{figure}

While HN3 is of a fixed, finite degree, there exist generalizations of
HN3 that lead to new, revealing insights into small-world
phenomena~\citep{SWPRL,SWN,Boettcher11b}. For instance, we can extend
HN3 in the following manner to obtain a network of average degree
5, hence called HN5. In addition to the edges in HN3, in HN5 we also
connect each site in level $i$ ($i\geq2$, i.e., all even sites), to
(higher-level) sites that are  $2^{i-1}$ sites away in both
  directions. Note that Eq.~(\ref{eq:numbering}) implies that the
nearest neighbors of a site $i$ within its hierarchy are separated by a distance
of $2\times 2^{i-1}$.  The resulting network HN5 remains planar but
now sites have a hierarchy-dependent degree, as shown in
Fig.~\ref{fig:5hanoi}. To obtain the average degree, we observe that
1/2 of all sites have degree 3, 1/4 have degree 5, 1/8 have degree 7,
and so on, leading to an exponentially falling degree distribution of
${\cal P}\left\{ \alpha=2i+1\right\} \propto2^{-i}$. Then, the total
number of edges $L$ in a system of size $N=2^{k}+1$ as shown in
Fig.~\ref{fig:5hanoi} is
\begin{eqnarray}
2L & =2\left(2k+1\right)+ & \sum_{i=1}^{k-1}\left(2i+1\right)2^{k-i}=5\times2^{k}-4,
\label{eq:TotalLinksHN5}
\end{eqnarray}
where the expression outside the sum refers to the special case of
those three vertices at the highest levels, $k-1$ and $k$. Any other
choice of boundary conditions may vary the offset in
Eq.~(\ref{eq:TotalLinksHN5}), but not the average degree, which is
\begin{eqnarray}
\left\langle \alpha\right\rangle  & = & \frac{2L}{N}\sim5.
\label{eq:averageDegreeHN5}
\end{eqnarray}

In HN5, the end-to-end distance is trivially 1 (see
Fig.~\ref{fig:5hanoi}).  Therefore, we define as the diameter the
largest of the shortest paths possible between any two sites, which
are typically odd-index sites farthest away from long-distance
edges. For the $N=33$ site network depicted in Fig.~\ref{fig:5hanoi},
for instance, that diameter is 5, measured between sites 3 and 19
(starting with $n=0$ as the left-most site), although there are many
other such pairs. It is easy to show recursively that this diameter
grows as
\begin{eqnarray}
d^{\rm HN5} & = & 2\left\lfloor k/2\right\rfloor +1\sim\log_{2}N.
\label{eq:5dia}
\end{eqnarray}
Other variants of the Hanoi networks are conceivable. For instance, a
non-planar version has been
designed~\citep{Boettcher09c,Boettcher10c}, but that network happens
to possess only a unique, alternating covering of $\frac{1}{2}$ and is
not considered here.

\section{Vertex-Cover Problem as a Hard-Core Lattice Gas\label{sec:Vertex-Cover-Problem}}

Vertex cover is a well-known NP-hard combinatorial
problem~\citep{Karp72,G+J,Ausiello99} that consists of finding a
minimal covering of the vertices of a network in such a way
that each edge is covered at least once.  Formally, for a graph
$G=(V,E)$, with $V$ being the set of vertices and $E\subset V^{(2)}$ 
the set of edges, a vertex cover $V'$ is a subset of $V$ with the
property that for each (undirected) edge $\{i,\,j\}\in E$ either $i\in
V'$ or $j\in V'$.  A minimum vertex cover $V_{\min}$ is a vertex cover
of minimum cardinality $|V_{\min}|$.

As shown in Ref.~\citep{Weigt01}, the vertex-cover problem can be formulated
alternatively as a hard-core repulsive lattice gas problem. In this
formulation, the uncovered vertices of the covering problems
correspond to the actual gas particles. These particles have a
hard-core repulsion such that they can not occupy neighboring lattice
sites, i.e., they cannot simultaneously vie for the same
edge. Interpreting these particles as the voids of the covering
problem implies that no edge may be left uncovered on both
ends. Accordingly, all properties of the minimum cover problem derive
from the ground state of the lattice gas at its highest
packing.

The grand canonical partition function for such a lattice gas is
generically given by
\begin{eqnarray}
\Xi\left(\mu\right) & = &
\sum_{x_{0}=\{0,1\}}\ldots\sum_{x_{N}=\{0,1\}}\exp\left(
\mu\sum_{i=1}^{N}x_{i}\right) \prod_{\langle i,j\rangle}\left(1-x_{i}x_{j}\right),
\label{eq:GrandPF}
\end{eqnarray}
where the product extends over all edges of the graph and exerts the
hard-core repulsive constraint. The chemical potential $\mu$ is
provided to regulate the density as gas particles get packed into the
system.  Since maximal density of the gas implies minimal coverage of
all edges, we are looking for the configurations in the limit
$\mu\to\infty$ of the gas.

The quantities~\citep{Weigt01} we seek are the thermodynamic limit
($N\to\infty$) of the packing fraction for the lattice gas,
\begin{eqnarray}
\nu\left(\mu\right) & = & \frac{1}{N}\left\langle \sum_{i=1}^{N}x_{i}\right\rangle _{\mu}=\frac{1}{N}\frac{\partial}{\partial\mu}\ln\Xi\left(\mu\right),
\label{eq:nu}
\end{eqnarray}
and the entropy density of such configurations,
\begin{eqnarray}
s\left(\nu\left(\mu\right)\right) & = & \frac{1}{N}\left(1-\mu\frac{\partial}{\partial\mu}\right)\ln\Xi\left(\mu\right).
\label{eq:entropy}
\end{eqnarray}
It has also been shown in Ref.~\citep{Weigt01} that one can extract the
corresponding properties of the minimal vertex coverage from these
in the $\mu\to\infty$ limit. For the coverage density, this corresponds
simply to the void density of the gas,
\begin{eqnarray}
c_{\rm min} & = & 1-\lim_{\mu\to\infty}\nu\left(\mu\right),
\label{eq:VCnu}
\end{eqnarray}
and the entropy density of optimal coverages is simply equal to that
for the lattice gas: 
\begin{eqnarray}
s_{\rm VC}\left(c_{\rm min}\right) & = & s\left(\nu=1-c_{\rm min}\right).
\label{eq:VCentropy}
\end{eqnarray}
Due to the hierarchical structure of the Hanoi networks, we will also
introduce level-specific chemical potentials $\mu_{i}$, for example,
to extract information about the coverage with respect to the level of
the hierarchy (i.e., the range its small-world edge attains) that a
vertex may reside in. The corresponding derivations are presented in
the Appendix. Throughout, we will find it often convenient to express
the chemical potentials as an activity variable,
\begin{eqnarray}
m_{i} & = & e^{-\mu_{i}}\qquad\left(1\leq i\leq k\right),
\label{eq:Defm}
\end{eqnarray}
such that $\mu_{i}\to\infty$ corresponds to the somewhat more tractable
limit $m_{i}\to0$.

\section{RG for the Hard-Core Lattice Gas on Hanoi Networks\label{sec:RG-for-the}}

The renormalization group (RG) as applied to the lattice-gas problem
developed here contains a few unfamiliar features. Thus, we have to elaborate
to a significant extend on the procedure. Although ultimately the RG will
 heavily rely on procedures used for Ising spin models,
initially we will have to rewrite the grand canonical partition
function of the lattice gas in an appropriate form. To this end, the
purpose of the first step of the RG (already eliminating half of all
sites) is to generate the initial conditions for the subsequent
canonical partition function analysis, in which the usual coupling
variables depend in a complicated way on the chemical potential $\mu$
instead of a temperature, and the apparent {}``spin'' variables are in
fact Boolean, $x_{i}\in\left\{ 0,1\right\} $.

We have to rewrite the generic partition function in
Eq.~(\ref{eq:GrandPF}) for the special case of the Hanoi networks. To
access more details of the solutions, we will take the opportunity to
generalize to the case of a hierarchy-specific chemical potential
$\mu_{i}$ for $1\leq i\leq k$, where $N=2^{k}+1$ is the size of the
system. (For the RG, it is natural to consider the Hanoi network with
an open boundary both at node $0$ and at node $2^{k}$; for a system
with periodic boundaries on a loop, both of these nodes would become
identical and $N=2^{k}$ would be the size of the system. Of course,
either choice results in identical thermodynamic averages.)

First, we rewrite the hard-core repulsive factor in
Eq.~(\ref{eq:GrandPF}) as separate products, one for the long-range
edges and the other for the backbone edges,
\begin{eqnarray}
\prod_{\langle i,j\rangle}\left(1-x_{i}x_{j}\right) & = & \left[\prod_{i=1}^{{\cal K}}\prod_{n=1}^{2^{k+1-i}}\left(1-x_{2^{i-1}\left(n-1\right)}x_{2^{i-1}n}\right)\right]\left[\prod_{i=1}^{k-1}\prod_{l=1}^{2^{k-i-1}}\left(1-x_{2^{i-1}\left(4l-3\right)}x_{2^{i-1}\left(4l-1\right)}\right)\right].
\label{eq:HardCoreProd}
\end{eqnarray}
The case ${\cal K}=1$ corresponds to HN3, with a simple,
one-dimensional line of edges connecting all sites in the backbone
sequentially. In turn, for HN5 we set ${\cal K}=k$,  with
each $i>1$ referring to the layers of those edges that connect along the
backbone only every second site, every fourth site, every eight site,
etc., as shown in Fig.~\ref{fig:5hanoi}. Note that in
Eq.~(\ref{eq:HardCoreProd}) we have used the decomposition of the
sites in the network implied by the renumbering in
Eq.~(\ref{eq:numbering}).

By the same token, we re-order the summation in Eq.~(\ref{eq:GrandPF})
as
\begin{eqnarray}
\sum_{x_{0}}e^{\mu_{i(0)}x_{0}}\ldots\sum_{x_{N}}e^{\mu_{i(N)}x_{N}} & = & \sum_{x_{0}}m_{i(0)}^{-x_{0}}\ldots\sum_{x_{N}}m_{i(N)}^{-x_{N}},
\label{eq:sum_renumber}\\
 & = &
\sum_{x_{0},x_{2^{k-1}},x_{2^{k}}}m_{k}^{-x_{0}-x_{2^{k-1}}-x_{2^{k}}}\left[\prod_{i=1}^{k-1}\prod_{l=1}^{2^{k-i-1}}\sum_{x_{2^{i-1}\left(4l-3\right)}}\sum_{x_{2^{i-1}\left(4l-1\right)}}m_{i}^{-x_{2^{i-1}\left(4l-3\right)}-x_{2^{i-1}\left(4l-1\right)}}\right],
\nonumber
\end{eqnarray}
where we have simplified the notation on the sums to mean
$\sum_{x}\hat{=}\sum_{x\in\left\{ 0,1\right\} }$.  Of course,
Eq.~(\ref{eq:sum_renumber}) has to be understood in an operator sense,
i.e., the summations extend to all site-variables that match the
indicated index. Here, we have also allowed for a site-specific
chemical potential. It is our goal to extract local packing
information, not for each site, but for all vertices within a specific
hierarchy, where $i(n)$ refers to the chemical potential in the $i$-th
level that the vertex $n$ is associated with according to
Eq.~(\ref{eq:numbering}).  Naturally, the sites at the highest level
$k$ of the hierarchy ($x_{0},x_{2^{k-1}},x_{2^{k}}$) require  special
consideration.

In this parametrization of the indices, the products in
Eq.~(\ref{eq:sum_renumber}) can be combined with those of the second
factor in Eq.~(\ref{eq:HardCoreProd}).  Both refer to the small-world
edges in all levels of the hierarchy and are naturally expressed in a
hierarchy-conform manner. Hence, we find for the grand-canonical
partition function defined in Eq.~(\ref{eq:GrandPF}) on a Hanoi
network with $k$ levels in the hierarchy:
\begin{eqnarray}
\Xi^{(k)}_{{\cal K}}\left(m_{1,}\dots,m_{k}\right) & = & \sum_{x_{0},x_{2^{k-1}},x_{2^{k}}}m_{k}^{-x_{0}-x_{2^{k-1}}-x_{2^{k}}}\,{\cal S}_{{\cal K}}\left(m_{2,}\dots,m_{k-1}\right)\,\prod_{j=1}^{2^{k-2}}\Theta\left(m_{1},x_{2\left(2j-2\right)},x_{2\left(2j-1\right)},x_{2\left(2j\right)}\right),
\label{eq:badPF-1}
\end{eqnarray}
where we have defined the operator for the weighted summation on HN3
and HN5, respectively,
\begin{eqnarray*}
{\cal S}_{\rm HN3} & \equiv & \prod_{i=2}^{k-1}\prod_{l=1}^{2^{k-i-1}}\sum_{x_{2^{i-1}\left(4l-3\right)}}\sum_{x_{2^{i-1}\left(4l-1\right)}}m_{i}^{-x_{2^{i-1}\left(4l-3\right)}-x_{2^{i-1}\left(4l-1\right)}}\left(1-x_{2^{i-1}\left(4l-3\right)}x_{2^{i-1}\left(4l-1\right)}\right),\\
{\cal S}_{\rm HN5} & \equiv &
\prod_{i=2}^{k-1}\prod_{l=1}^{2^{k-i-1}}\sum_{x_{2^{i-1}\left(4l-3\right)}}\sum_{x_{2^{i-1}\left(4l-1\right)}}m_{i}^{-x_{2^{i-1}\left(4l-3\right)}-x_{2^{i-1}\left(4l-1\right)}}\left(1-x_{2^{i-1}\left(4l-3\right)}x_{2^{i-1}\left(4l-1\right)}\right)
\nonumber\\
 &  & \times\left(1-x_{2^{i-1}\left(4l-4\right)}x_{2^{i-1}\left(4l-3\right)}\right)\left(1-x_{2^{i-1}\left(4l-3\right)}x_{2^{i-1}\left(4l-2\right)}\right)\left(1-x_{2^{i-1}\left(4l-2\right)}x_{2^{i-1}\left(4l-1\right)}\right)\left(1-x_{2^{i-1}\left(4l-1\right)}x_{2^{i-1}\left(4l\right)}\right).\nonumber
\end{eqnarray*}
Note that these operators only sum over all even-indexed variables
(i.e., $i\geq2$). To obtain a renormalizable form for the partition
function it is necessary to trace over the lowest level $i=1$ of
the hierarchy, i.e., to eliminate all odd-index variables. For both,
HN3 and HN5, this results in an identical structure, defined as 
\begin{eqnarray}
\label{eq:Theta}
\Theta\left(\mu_{1},x_{2\left(2j-2\right)},x_{2\left(2j-1\right)},x_{2\left(2j\right)}\right) & = & \sum_{x_{4j-3}}\sum_{x_{4j-1}}m_{1}^{-x_{4l-3}-x_{4l-1}}\left(1-x_{4j-3}x_{4j-1}\right)\\
 &  & \quad\left(1-x_{4j-4}x_{4j-3}\right)\left(1-x_{4j-3}x_{4j-2}\right)\left(1-x_{4j-2}x_{4j-1}\right)\left(1-x_{4j-1}x_{4j}\right),
\nonumber \\
 & = & 1+e^{\mu_{1}}\left(1-x_{2\left(2j-1\right)}\right)\left(2-x_{2\left(2j-2\right)}-x_{2\left(2j\right)}\right).
\nonumber 
\end{eqnarray}
In Appendix~A, we show how to
recast $\Theta$ in an Ising-like form with a sufficient number of
renormalizable parameters. We can simplify the grand partition
function in Eq.~(\ref{eq:badPF-1}) further by combining the products
and writing
\begin{eqnarray}
\Xi^{(k)}\left(m_{1},\ldots,m_{k}\right) & = & \sum_{x_{0},x_{2^{k-1}},x_{2^{k}}}m_{k}^{-x_{0}-x_{2^{k-1}}-x_{2^{k}}}\left[\prod_{i=2}^{k-2}\prod_{l=1}^{2^{k-i-2}}\sum_{x_{2^{i}\left(4l-3\right)}}\sum_{x_{2^{i}\left(4l-1\right)}}\right]\prod_{l=1}^{2^{k-3}}\zeta_{1}^{l}\left(x_{4\left(2l-2\right)},x_{4\left(2l-1\right)},x_{4\left(2l\right)}\right),
\label{eq:goodPF}
\end{eqnarray}
where the explicit expression for $\zeta_{1}^{l}$ is also derived in
Appendix~A for both, HN3 and HN5,
which allows us to drop the subscript label. In either case, the RG
recursion equations now result from imposing the recursive relation
between hierarchies,
\begin{eqnarray}
 &  & \zeta_{i+1}^{l}\left(x_{2^{i+1}\left(2l-2\right)},x_{2^{i+1}\left(2l-1\right)},x_{2^{i+1}\left(2l\right)}\right)
 \label{eq:zeta_recur}\\
 & = &
 \sum_{x_{2^{i}\left(4l-3\right)}}\sum_{x_{2^{i}\left(4l-1\right)}}\zeta_{i}^{2l-1}\left(x_{2^{i}\left(4l-4\right)},x_{2^{i}\left(4l-3\right)},x_{2^{i}\left(4l-2\right)}\right)\zeta_{i}^{2l}\left(x_{2^{i}\left(4l-2\right)},x_{2^{i}\left(4l-1\right)},x_{2^{i}\left(4l\right)}\right),
\nonumber
\end{eqnarray}
which are derived in Appendix~A.
There, Figs.~\ref{fig:RG3} and \ref{fig:RG5} also provide a graphical
representation of Eq.~(\ref{eq:zeta_recur}).

\subsection{Analysis of the RG Recursions\label{sub:AnalysisRG}}

We find that the RG recursions that follow from the previous
discussion,  which are given explicitly in
Eqs.~(\ref{eq:RGrecurHN3}) for HN3 and in Eqs.~(\ref{eq:RGrecurHN5})
for HN5 for the hard-core lattice gas model,  have only two trivial
fixed points. There is a stable low-density fixed point for all $\mu<\infty$,
i.e., $m>0$, and an unstable fixed point at full-packing for
$\mu=\infty$, i.e., $m=0$. Note that in this part of the analysis we
are concerned with global properties, and thus, ignore differences
between the hierarchical level by setting $m_{i}\equiv m$ throughout.

\subsubsection{Analysis for HN3\label{subsub:Analysis-for-HN3}}

The limit $m\to0$ of the recursions in Eqs.~(\ref{eq:RGrecurHN3}) for
initial conditions given in Eqs.~(\ref{eq:RGIC}) is difficult to handle. Except
for $\kappa_{1}$, all other parameters are either diverging or
vanishing in Eqs.~(\ref{eq:RGIC2}) for that limit. To achieve a
clearer picture, we evolve the recursions once and obtain
\begin{eqnarray}
\eta_{2}\sim\frac{24}{5},\quad\gamma_{2}\sim\frac{8}{3},\quad C_{2}\sim\frac{m^{2}}{8}, &  & \kappa_{2}\sim\frac{15}{8},\quad\lambda_{2}\sim\frac{25}{24},\quad\Delta_{2}\sim\frac{4}{25m}.
\label{eq:RGIC2}
\end{eqnarray}
In fact, further revolutions in the recursions seems to preserve this
picture: $C_{i}$ scales with a rapidly growing power of $m$, while all
other parameters and $\bar{\Delta}_{i}=m\Delta_{i}$ become finite for
$m=0$ at any order $i$. Thus, we replace $\Delta$ with $\bar{\Delta}$
and subsequently set $m\to0$ in Eqs.~(\ref{eq:RGrecurHN3}) yielding
\begin{eqnarray}
C_{i+1}  \sim  \frac{m\gamma_{i}C_{i}^{2}}{2},
&&\gamma_{i+1}  \sim  \gamma_{i}\eta_{i}\kappa_{i},\qquad
\eta_{i+1}  \sim  \frac{4\kappa_{i}}{\left(1+\kappa_{i}\right)^{2}},
\nonumber \\
\kappa_{i+1}  \sim  \lambda_{i}\frac{\left(1+\kappa_{i}\right)}{\kappa_{i}},
&&\lambda_{i+1} \sim  \frac{\left(1+\kappa_{i}\right)^{2}}{4\kappa_{i}},\qquad
\bar{\Delta}_{i+1}  \sim  \frac{2\kappa_{i}^{2}\bar{\Delta}_{i}}{\left(2+\gamma_{i}\kappa_{i}^{2}\bar{\Delta}_{i}\right)\left(1+\kappa_{i}\right)^{2}}.
\label{eq:FP_m0HN3} 
\end{eqnarray}
At its core, the two recursions for $\kappa$ and $\lambda$ have
become independent of all the others. The $m=0$ fixed-point itself
is then dominated solely by the stationary solution of their recursions
in Eqs.~(\ref{eq:FP_m0HN3}),
\begin{eqnarray}
\kappa^{*}=\frac{1}{2^{\frac{2}{3}}-1}, & \qquad & \lambda^{*}=\frac{1}{2^{\frac{2}{3}}\left(2^{\frac{2}{3}}-1\right)}.
\label{eq:FPHN3}
\end{eqnarray}
Therefore, one finds a constant solution for
$\eta^{*}=4\kappa^{*}/\left(1+\kappa^{*}\right)^{2}=1/\lambda^{*}$ and
the recursion
$\gamma_{i+1}\sim\gamma_{i}\left(\kappa^{*}/\lambda^{*}\right)$ with
the solution $\gamma_{i}\sim\gamma_{0}2^{\frac{2i}{3}}$ which diverges
for large $i$. The situation for $\bar{\Delta}_{i}$ is more
subtle. Numerics clearly indicates its decay, but this could occur
consistently in two ways. First, if it were to decay such that
$\gamma_{i}\bar{\Delta}_{i}$ still increases, then
Eq.~(\ref{eq:FP_m0HN3}) suggests
$\bar{\Delta}_{i+1}\propto1/\gamma_{i}$, but that would render
$\gamma_{i}\bar{\Delta}_{i}$ constant, which is a contradiction.
Alternatively, if both $\bar{\Delta}_{i}$ and
$\gamma_{i}\bar{\Delta}_{i}$ decay, then
$\bar{\Delta}_{i+1}\sim\bar{\Delta}_{i}\left[\kappa^{*}/\left(1+\kappa^{*}\right)\right]^{2}$,
yielding $\bar{\Delta}_{i}\sim2^{-\frac{4i}{3}}$ in a consistent
manner. Numerical studies verify that the latter solution is indeed
realized.

\begin{figure}
\includegraphics[bb=14bp 35bp 710bp 522bp,clip,scale=0.5]{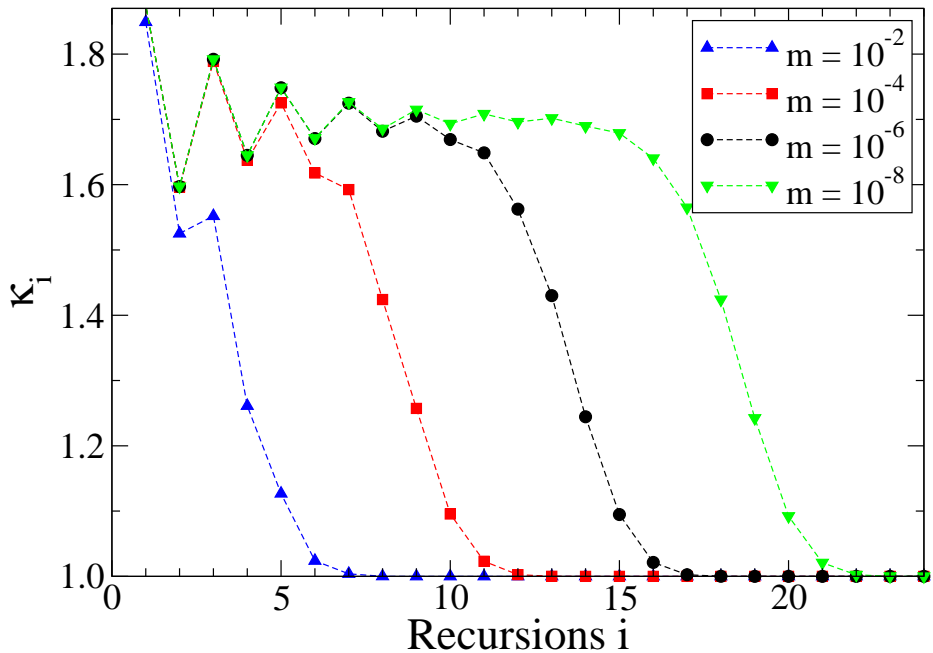}
\caption{\label{fig:xi_HN3}(Color online)
Plot of the value of $\kappa_{i}$ after the $i$-th RG-step for
$m=10^{-2},10^{-4},10^{-6},$ and $10^{-8}$ (left to right). At a
length scale $\xi\left(m\right)=2^{i}$ with $i=-\frac{3}{4}\log_{2}m$,
the behavior of $\kappa_{i}$ crosses over from the value at the
unstable $m=0$ fixed point,
$\kappa^{*}=1/\left(2^{2/3}-1\right)\approx1.70$, to the stable $m=1$
($\mu=0$) fixed point at which $\kappa^{*}=1$.}
\end{figure}

From the terms dropped in the $m\to0$ limit, we can extract a
cross-over scale as follows: Achieving the limit $m\to0$ implies that
the widely occurring term $m\gamma_{i}$ in Eqs.~(\ref{eq:RGrecurHN3})
is considered small enough to be discarded with respect to terms of
order unity.  Hence, by identifying $\xi=\sqrt{2^{i(m)}}$ as the
correlation length within the small-world metric supplied by
Eq.~(\ref{eq:3dia}), using $\gamma_{i(m)}\sim1/m$ yields $2^{i(m)}\sim
m^{-\frac{3}{2}}$ or
\begin{eqnarray}
\xi & \sim & e^{\frac{3}{4}\mu}
\label{eq:corrlength}
\end{eqnarray}
as the diverging length below which the systems orders for an
correspondingly diverging chemical potential, $\mu\to\infty$. Indeed,
 for $m=10^{-4}$, for example, we find numerically that the solution veers off
the unstable fixed point just below the $i=10$th iteration;
Fig.~\ref{fig:xi_HN3} demonstrates the correctness of
Eq.~(\ref{eq:corrlength}) for any small $m$.

\subsubsection{Analysis for HN5\label{subsub:Analysis-for-HN5}}

The analysis for HN5 is surprisingly subtle. Although the preceding fixed-point
analysis for HN3 required the singular limit $m\to0$ as part of
the consideration, after the appropriate rescaling of the parameters
with $m$, the subsequent approach proceeds in a familiar fashion.  HN5
obscures this approach with an additional layer of complexity,
resulting from strong alternating effects order-by-order in the RG, as
the numerics reveals. Of course, the initial conditions here are
identical to those for HN3 in Eqs.~(\ref{eq:RGIC}), with the same
pathologies in the $m\to0$ limit. However, whereas those problems were
essentially resolved for HN3 after one RG-step and rescaling, see Eqs.
(\ref{eq:RGIC2}), here we find
\begin{equation}
C_{2}  \sim \frac{m^{2}}{2},\quad\gamma_{2}\sim2,\quad\eta_{2}\sim\frac{8}{9},\quad
\kappa_{2}  \sim
\frac{3}{8m},\quad\lambda_{2}\sim\frac{9}{8},\quad\Delta_{2}\sim\frac{8}{9},
\label{eq:IC2HN5}
\end{equation}
and 
\begin{equation}
C_{3} \sim  \frac{m^{5}}{16},\quad\gamma_{3}\sim\frac{16}{9m},\quad\eta_{3}\sim16m,\quad
\kappa_{3}\sim\frac{9}{16},\quad\lambda_{3}\sim\frac{1}{16m},\quad\Delta_{3}\sim16m,
\label{eq:IC3HN5}
\end{equation}
etc. This alternation between regular and singular behaviors of each of
the parameters persists thereafter. Leaving the recursion for $C_{i}$
aside for now, we notice that for even indices, $\gamma_{2n}$,
$\eta_{2n}$, $m\kappa_{2n}$, $\lambda_{2n}$, and $\Delta_{2n}$ remain
finite for $m\to0$, but for odd indices, this is true for
$m\gamma_{2n-1}$, $\eta_{2n-1}/m$, $\kappa_{2n-1}$, $m\lambda_{2n-1}$,
and $\Delta_{2n-1}/m$.  Defining $\bar{\gamma}_{2n-1}=m\gamma_{2n-1}$,
$\bar{\eta}_{2n-1}=\eta_{2n-1}/m$, $\bar{\kappa}_{2n}=m\kappa_{2n}$,
$\bar{\lambda}_{2n-1}=m\lambda_{2n-1}$, and
$\bar{\Delta}_{2n-1}=\Delta_{2n-1}/m$, it is useful to rewrite the
recursions in Eqs.~(\ref{eq:RGrecurHN5}) separately for even and odd
indices. In fact, the limit $m\to0$ on its explicit appearance can now
be taken to get
\begin{eqnarray}
\gamma_{2n}=\bar{\eta}_{2n-1}\left(2+\bar{\gamma}_{2n-1}\right), & \quad & \bar{\gamma}_{2n-1}=\eta_{2\left(n-1\right)}\left(2+m\gamma_{2\left(n-1\right)}\right)\sim2\eta_{2\left(n-1\right)},\nonumber\\
\eta_{2n}=\bar{\gamma}_{2n-1}\frac{2+\bar{\gamma}_{2n-1}}{\left(1+\bar{\gamma}_{2n-1}\right)^{2}}, & \quad & \bar{\eta}_{2n-1}=\gamma_{2\left(n-1\right)}\frac{2+m\gamma_{2\left(n-1\right)}}{\left(1+m\gamma_{2\left(n-1\right)}\right)^{2}}\sim2\gamma_{2\left(n-1\right)},\nonumber\\
\bar{\kappa}_{2n}=\bar{\lambda}_{2n-1}\frac{\left(1+\bar{\gamma}_{2n-1}\right)^{2}}{2+\bar{\gamma}_{2n-1}}, & \quad & \kappa_{2n-1}=\lambda_{2\left(n-1\right)}\frac{1+m\gamma_{2\left(n-1\right)}}{2+m\gamma_{2\left(n-1\right)}}\sim\frac{1}{2}\lambda_{2\left(n-1\right)},\\
\lambda_{2n}=\frac{\left(1+\bar{\gamma}_{2n-1}\right)^{2}}{\bar{\gamma}_{2n-1}\left(2+\bar{\gamma}_{2n-1}\right)}, & \quad & \bar{\lambda}_{2n-1}=\frac{1+m\gamma_{2\left(n-1\right)}}{\gamma_{2\left(n-1\right)}\left(2+m\gamma_{2\left(n-1\right)}\right)}\sim\frac{1}{2\gamma_{2\left(n-1\right)}},\nonumber\\
\Delta_{2n}=\bar{\gamma}_{2n-1}\frac{2+\bar{\gamma}_{2n-1}}{\left(1+\bar{\gamma}_{2n-1}\right)^{2}}, & \quad & \bar{\Delta}_{2n-1}=\gamma_{2\left(n-1\right)}\frac{2+m\gamma_{2\left(n-1\right)}}{\left(1+m\gamma_{2\left(n-1\right)}\right)^{2}}\sim2\gamma_{2\left(n-1\right)}.\nonumber
\end{eqnarray}
Note that for the limit $m\to0$ we only assumed that $m\gamma_{2\left(n-1\right)}\ll1$
for $n\to\infty$ on the right-hand set of these relations, which
provides a correlation length from the cross-over $n_{\rm co}=n\left(m\right)$
at $\gamma_{2n_{\rm co}}\sim1/m$. Eliminating all odd-index quantities
from the equations yields
\begin{eqnarray}
\gamma_{2n}  =  4\gamma_{2\left(n-1\right)}\left(1+\eta_{2\left(n-1\right)}\right),&&
\eta_{2n}  = 4\eta_{2\left(n-1\right)}\frac{1+\eta_{2\left(n-1\right)}}{\left(1+2\eta_{2\left(n-1\right)}\right)^{2}},\\
\bar{\kappa}_{2n}  =  \frac{1+2\eta_{2\left(n-1\right)}}{4\gamma_{2\left(n-1\right)}\left(1+\eta_{2\left(n-1\right)}\right)},&&
\lambda_{2n}  =  \frac{\left(1+2\eta_{2\left(n-1\right)}\right)^{2}}{4\eta_{2\left(n-1\right)}\left(1+\eta_{2\left(n-1\right)}\right)},\qquad
\Delta_{2n}  =  4\eta_{2\left(n-1\right)}\frac{1+\eta_{2\left(n-1\right)}}{\left(1+2\eta_{2\left(n-1\right)}\right)^{2}}.\nonumber
\end{eqnarray}
These interlacing recursions now have a simple fixed point,
which derives from the only non-trivial solution of the self-contained
$\eta$-equation:
\begin{equation}
\eta^{*}  =  \frac{\sqrt{3}}{2}.
\end{equation}
This implies the equally stationary value
\begin{equation}
\Delta^{*}  =  \frac{1}{\lambda^{*}}=\frac{4\eta^{*}\left(1+\eta^{*}\right)}{\left(1+2\eta^{*}\right)}=\frac{3+\sqrt{3}}{2},
\end{equation}
but we also find the asymptotically scaling
\begin{equation}
\gamma_{2n}  \sim  \gamma_{0}\left[2\left(2+\sqrt{3}\right)\right]^{n}\propto\frac{1}{\bar{\kappa}_{2n}}.
\end{equation}
This provides the correlation length estimate
\begin{equation}
\xi  =  2^{n_{\rm co}}\sim\exp\left\{ \frac{\mu}{\log_{2}\left[2\left(2+\sqrt{3}\right)\right]}\right\} .
\end{equation}

\subsection{Packing Fraction and Entropy\label{sub:Coverage-and-Entropy}}

To understand the most pertinent features of the problem, such as the
optimal packing (or coverage) and its entropy, we have to consider the
asymptotic behavior of the renormalization group parameter $C_{i}$,
related to the growth of the overall energy-scale, in
Eq.~(\ref{eq:FP_m0HN3}) for the initial condition in
Eq.~(\ref{eq:RGIC2}). Clearly, the partition function at any finite
system size is a polynomial in $e^{\mu}$, i.e., in powers of
$m^{-1}$. Both of these quantities, packing fraction and entropy, derive from
the most divergent power in $m$ to be found in $\Xi$. To wit, we can
write for $m\to0$ with $N=2^k+1$,
\begin{eqnarray}
\Xi^{(k)} & \sim & \left(\sigma m^{-\alpha}\right)^{N}\left[1+am+bm^{2}+\ldots\right].
\label{eq:Xiasymp}
\end{eqnarray}
Then, it is $\partial_{\mu}\ln\Xi=-m\partial_{m}\ln\Xi\sim N\alpha$,
and we find from Eqs.~(\ref{eq:nu} and \ref{eq:entropy}),
\begin{eqnarray*}
\nu & = & \alpha,\\
s & = & \ln\sigma,
\end{eqnarray*}
for $N\to\infty$ at $m=0$. 

Equation~(\ref{eq:goodPF}) provides the grand canonical partition function
$\Xi^{(k)}$ for $2^{k}+1$ site-occupation variables in terms of an
Ising-like canonical partition function ${\cal Z}^{(k-1)}$ for only
$2^{k-1}+1$ (Boolean) spin variables. While $\Xi^{(k)}$  depends only on
the hierarchical chemical potentials $m_{i}$, ostensibly ${\cal
  Z}^{(k-1)}$ depends on a tuple $\vec{A}_{1}$ of renormalizable
couplings, see Eq.~(\ref{eq:Avector}), in addition to any
explicit dependence on $m_{i}$. Of course, the couplings themselves
are merely a function of the chemical potentials,
$\vec{A}_{1}=\vec{A}_{1}\left(m_{1}\right)$, through the RG initial
conditions in Eq.~(\ref{eq:RGIC}). Step by step in the RG, the
couplings transform according to Eq.~(\ref{eq:RGA}) each time the
system size halves, whereas the partition function stays
invariant. Hence, we can expand on Eq.~(\ref{eq:goodPF}) and write
\begin{eqnarray}
\Xi^{(k)}\left(m_{1},\ldots,m_{k}\right) & = & {\cal Z}^{(k-1)}\left(\vec{A}_{1}\left(m_{1}\right),m_{2},\ldots,m_{k}\right),
 \nonumber\\
 & = & {\cal Z}^{(k-2)}\left(\vec{A}_{2}\left(m_{1},m_{2}\right),m_{3},\ldots,m_{k}\right),
 \nonumber \\
 & \vdots &  \label{eq:goodPFchain}\\
 & = & {\cal Z}^{(1)}\left(\vec{A}_{k-1}\left(m_{1},\ldots,m_{k-1}\right),m_{k}\right), 
\nonumber 
 \end{eqnarray}
where ${\cal Z}^{(1)}$ is simply a rudimentary Hanoi network consisting
of just three vertices.

\subsubsection{Results for HN3\label{subsub:Results-for-HN3}}

\begin{figure}
\includegraphics[clip]{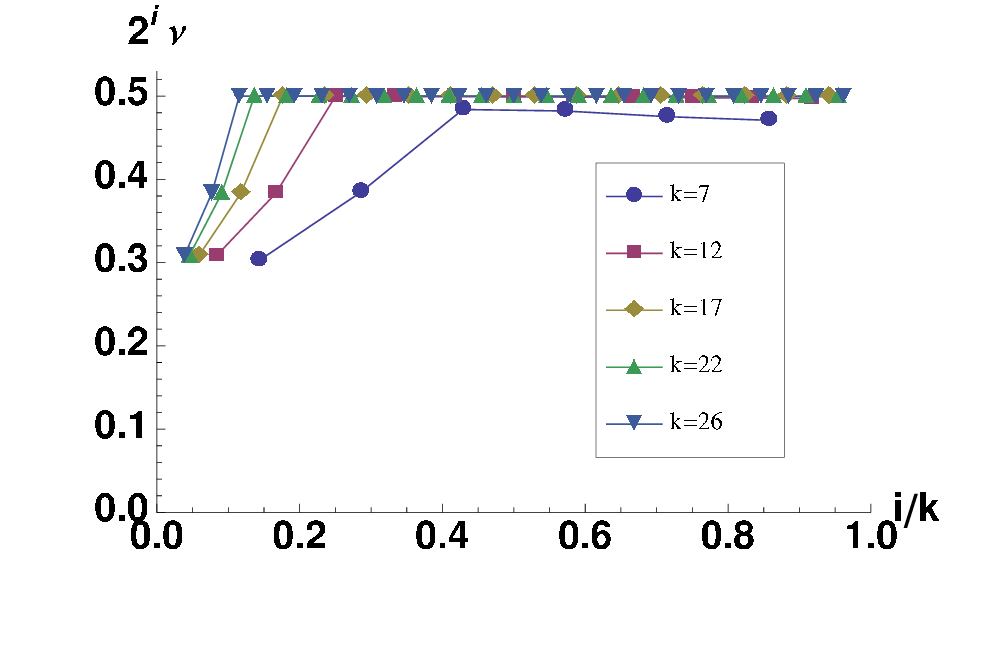}
\caption{\label{fig:levelnu_HN3}
Plot of the relative packing per level $2^{i}\nu_{i}$ on HN3 for
various system sizes $N=2^{k}+1$ with $k=7,12,17,22$, and 26, plotted
also on a relative level-scale $i/k$ at $m\to0$. Asymptotically, in
large systems, all vertices in higher levels $i$ appear to be just
50\% packed (or covered), which is minimally necessary to cover the
one small-world edge connecting such vertices. (Of course, each level
contains half as many vertices as any preceding level and thus contributes
ever less to the overall coverage.) This packing may well be random as
such vertices are far separated between the higher levels. A
significantly lower packing (higher coverage) is attained only at an
ever small fraction of the lowest levels to account for the overall
packing fraction of $\frac{3}{8}$ (coverage $\frac{5}{8}$). }
\end{figure}

Specializing this discussion for HN3, we find for the rudimentary
partition function ${\cal Z}^{(1)}$ in this case 
\begin{eqnarray}
{\cal Z}^{(1)} & = & C_{k-1}^{-1}\sum_{x_{0}}\sum_{x_{2^{k-1}}}\sum_{x_{2^{k}}}m_{k}^{-\left(x_{0}+x_{2^{k-1}}+x_{2^{k}}\right)}\gamma_{k-1}^{-\frac{1}{2}\left[\left(x_{0}+x_{2^{k-1}}\right)+\left(x_{2^{k-1}}+x_{2^{k}}\right)\right]}
\nonumber \\
 &  & \qquad\eta_{k-1}^{-\frac{1}{2}\left(x_{0}+x_{2^{k}}\right)}\kappa_{k-1}^{-\left(x_{0}x_{2^{k-1}}+x_{2^{k-1}}x_{2^{k}}\right)}\lambda_{k-1}^{-x_{0}x_{2^{k}}}\Delta_{k-1}^{-x_{0}x_{2^{k-1}}x_{2^{k}}}.
 \label{eq:Z1}
 \end{eqnarray}
For a uniform chemical potential, $m_{i}\equiv m$ for all $i$, one finds
that for $m\to0$ the partition function is dominated overwhelmingly
by the renormalized value of $C_{i}$, i.e.
\begin{eqnarray}
\ln\Xi^{(k)}\left(\mu\right) & = & \ln{\cal Z}^{(1)}(\vec{A}_{k-1}\left(m\right),m)\sim-\ln C_{k-1}.
\label{eq:lnC}
\end{eqnarray}
Rewriting the recursion for $C_{i}$ in Eq.~(\ref{eq:FP_m0HN3}) in
this form yields
\begin{eqnarray}
\ln C_{i+1} & = & 2\ln C_{i}+\ln\left(\frac{m\gamma_{i}}{2}\right)\sim2\ln C_{i}+\frac{2i}{3}\ln2+\ln\left(\frac{m\gamma_{0}}{2}\right),
\label{eq:lnCrecur}
\end{eqnarray}
which is easily summed up to give
\begin{eqnarray}
\ln C_{k-1} & = & 2^{k-3}\left[\ln C_{2}+\ln\left(2m\gamma_{0}\right)\right].
\label{eq:lnCk-1}
\end{eqnarray}
With $C_{2}\sim m^{2}$, as listed in Eq.~(\ref{eq:RGIC2}), we get
\begin{eqnarray}
\frac{1}{2^{k}}\ln\Xi^{(k)} & \sim & -\frac{1}{2^{k}}\ln C_{k-1}\sim-\frac{3}{8}\ln\left(m\right)-\frac{1}{8}\ln\left(4\gamma_{0}\right),
\label{eq:lnXik}
\end{eqnarray}
and comparison with Eq.~(\ref{eq:Xiasymp}) produces an exact prediction
for the maximal packing fraction of the lattice gas,
\begin{eqnarray}
\nu\left(\mu\to\infty\right) & = & \frac{3}{8},
\label{eq:nuHN3}
\end{eqnarray}
i.e., for the minimal fraction of vertices needing cover in HN3, it
is
\begin{eqnarray}
c_{\rm min} & = & \frac{5}{8}.
\label{eq:mincover58}
\end{eqnarray}
Note that the $m$-dependence of $C_{2}$ and of the recursion for
$C_{i}$ in Eqs.~(\ref{eq:FP_m0HN3}) are crucial for this result,
whereas $\gamma_{i}$ is independent of $m$ and, hence, becomes
irrelevant here. Unfortunately, the entropy density in turn depends
not only on the asymptotic form for $\gamma_{i}$ but on the
non-trivial integration constant $\gamma_{0}$, which can not be
determined from the asymptotic behavior of the RG flow; it is a
global property of that flow and could depend on all its
details. However, the result suggest that, at least for HN3, unlike for
those lattices in Fig.~\ref{fig:SierpinskiVC}, the entropy
density does not vanish but attains a non-trivial value. In fact,
using the recursions in Eqs.~(\ref{eq:RGrecurHN3}) for arbitrary $m$
and taking the $m\to0$ limit only in the end, we can exactly determine
the constant $\sigma$ defined in Eq.~(\ref{eq:Xiasymp}) for the first
few values of $k$ (see Tab.~\ref{tab:sigma}). Finite-size
extrapolation from the numerical evolution of the RG flow up to $k=25$
levels (i.e., system size $N=2^{25}$) for a finite but small value of
$m=10^{-40}$ predicts that $s_{\rm VC}(c_{\rm min})=0.160426(1)$.  (Any
variation of $m$ over 10 decades does not affect the extrapolation at
this accuracy.) For smaller system sizes we plot the packing fraction and the
entropy density for the entire range of the chemical potential in
Fig.~\ref{fig:entropy_HN3}.  In
Appendix, we describe how to
evaluate derivatives of the partition function, such as those leading
to $\nu$ and $s$, within the RG-scheme. There we also develop a
method to probe the packing fraction for each level of the
hierarchy; those results are plotted in Fig.~\ref{fig:levelnu_HN3}.

\begin{figure}
\hfil\includegraphics[clip,scale=0.8]{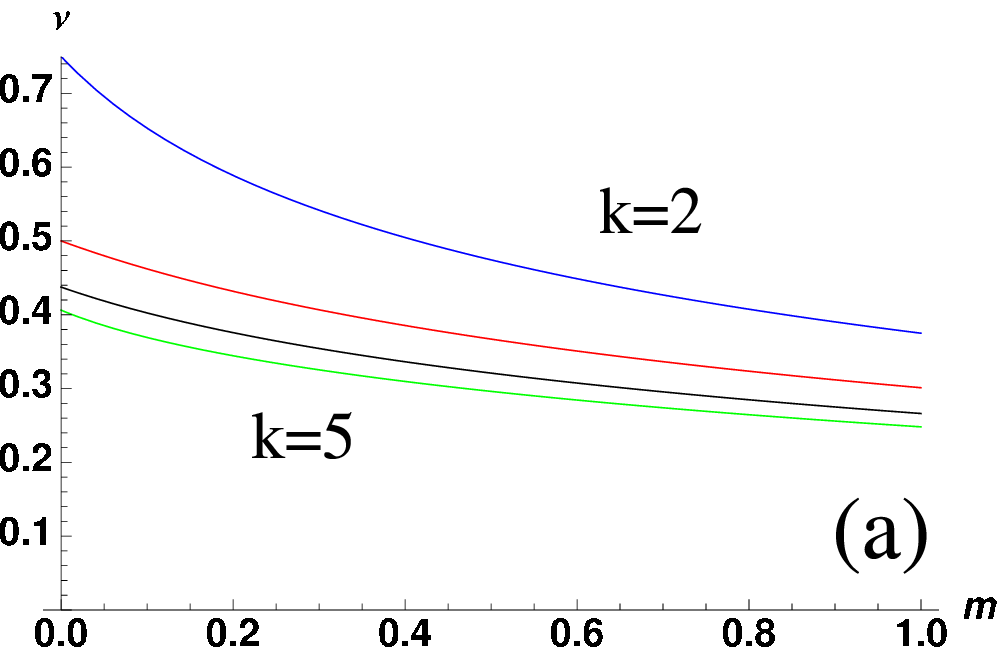}\hfil\includegraphics[clip,scale=0.8]{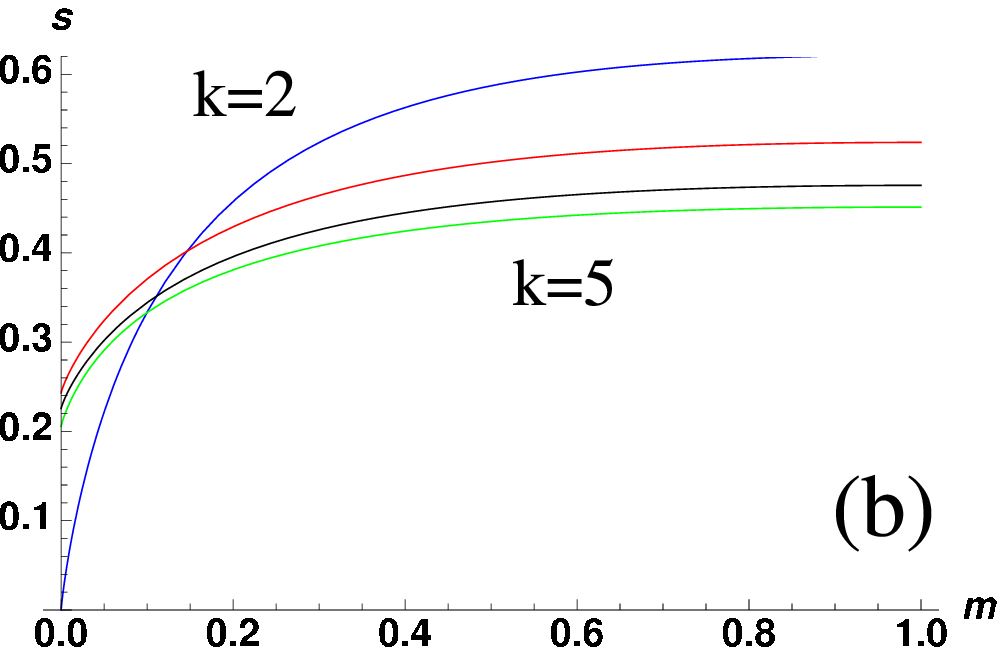}\hfil
\caption{\label{fig:entropy_HN3}(Color online)
Plot of (a) the packing fraction $\nu_{\rm VC}$  and (b) its entropy density
$s_{\rm VC}$  for the lattice-gas problem on HN3 for the first few
system sizes $N=2^{k}+1$ with $k=2,\ldots,5$ (top to bottom at $m=1$)
as a function of $m$.}
\end{figure}

\begin{table}
\caption{\label{tab:sigma}
Listing of the first few values of $\sigma$ and $s_{\rm VC}$ defined in
Eqs.~(\ref{eq:Xiasymp}) and (\ref{eq:VCentropy}) for HN3 of size
$N=2^{k}+1$. The sequence for the total number of optimal
configurations, $\sigma^{N}$, soon develops non-trivial prime
factors. The entropy density for the coverage $s_{\rm VC}$ only 
converges slowly to its numerical limit.}
\begin{tabular}{|c|r|l|}
\hline 
$k$ & $\sigma^{N}\qquad$ & $s_{\rm VC}=\ln\sigma$\tabularnewline
\hline
2 & 1 & 0\tabularnewline
3 & 7 & 0.243239\tabularnewline
4 & 37 & 0.225682\tabularnewline
5 & 718 & 0.205515\tabularnewline
6 & 193284 & 0.190186\tabularnewline
7 & 8651040480 & 0.178757\tabularnewline
8 & 11491993035377280000 & 0.171438\tabularnewline
$\vdots$ &$\vdots$  & $\vdots$\tabularnewline
$\infty$ &  & 0.160426(1)\tabularnewline
\hline
\end{tabular}
\end{table}

\begin{figure}
\hfill{}\includegraphics[angle=-90,scale=0.38]{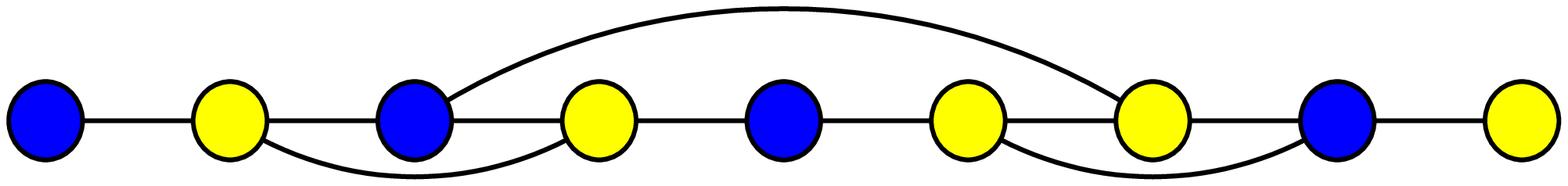}\hfill{}\includegraphics[angle=-90,scale=0.38]{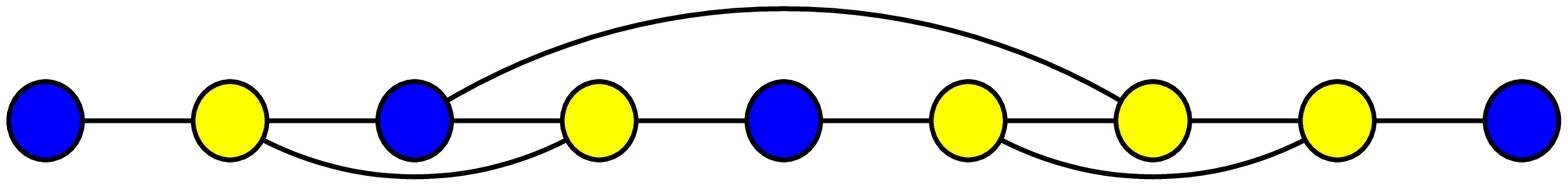}\hfill{}

\hfill{}\includegraphics[angle=-90,scale=0.38]{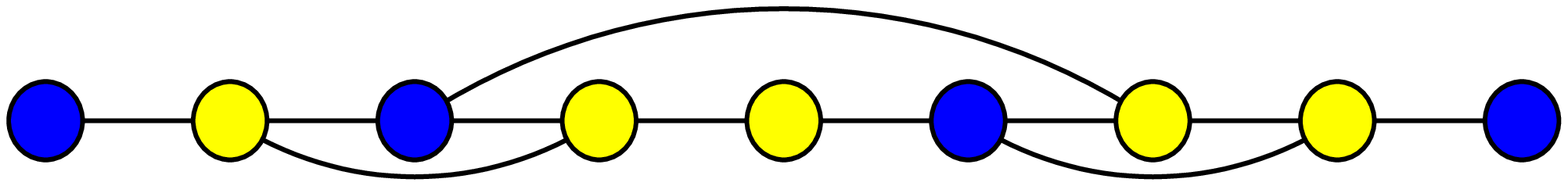}\hfill{}\includegraphics[angle=-90,scale=0.38]{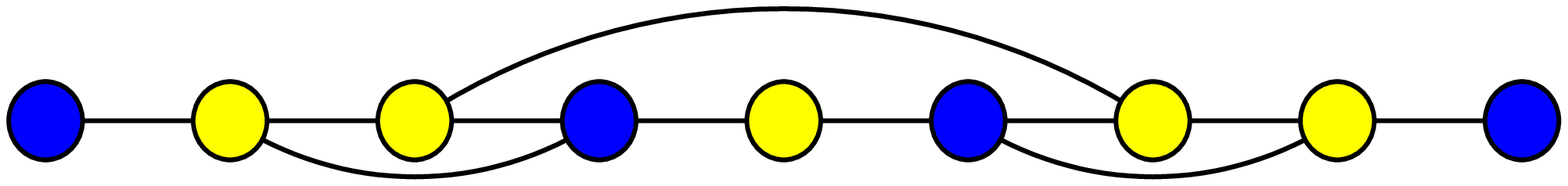}\hfill{}

\caption{\label{fig:Depiction_k3}
Depiction of perfect coverings on HN3 for $k=3$. Of all seven
solutions, we omitted the three obtained by reflection from
these. Light-colored sites belong to the vertex cover, dark-colored
sites mark particles with hard-core repulsion that prevents
nearest-neighbor occupation.}
\end{figure}

\begin{figure}
\hfill{}\includegraphics[angle=-90,scale=0.4]{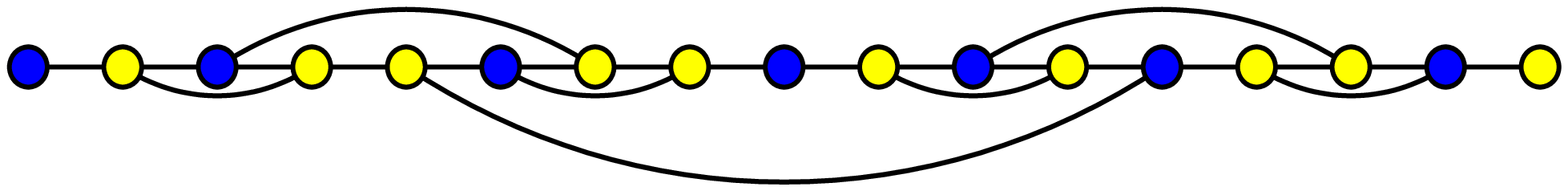}\hfill{}\includegraphics[angle=-90,scale=0.4]{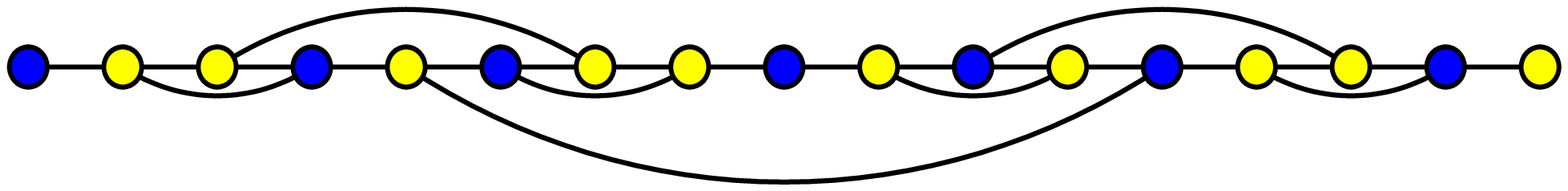}\hfill{}

\hfill{}\includegraphics[angle=-90,scale=0.4]{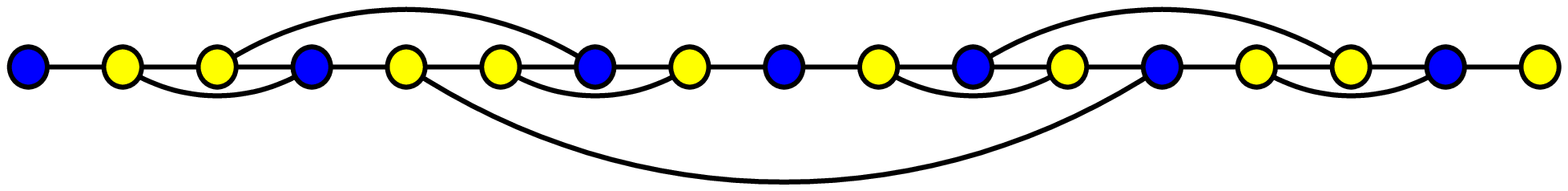}\hfill{}\includegraphics[angle=-90,scale=0.4]{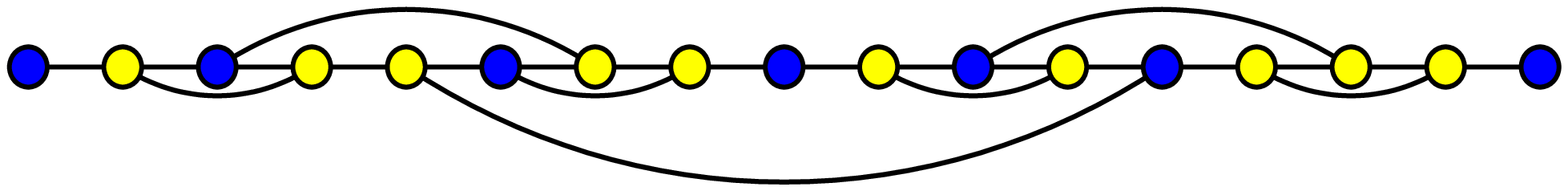}\hfill{}

\hfill{}\includegraphics[angle=-90,scale=0.4]{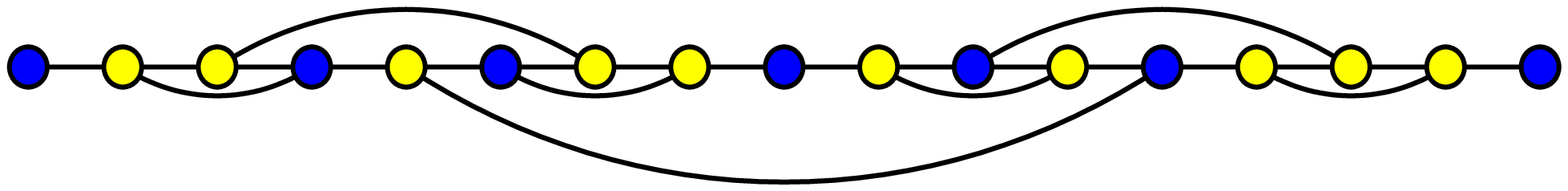}\hfill{}\includegraphics[angle=-90,scale=0.4]{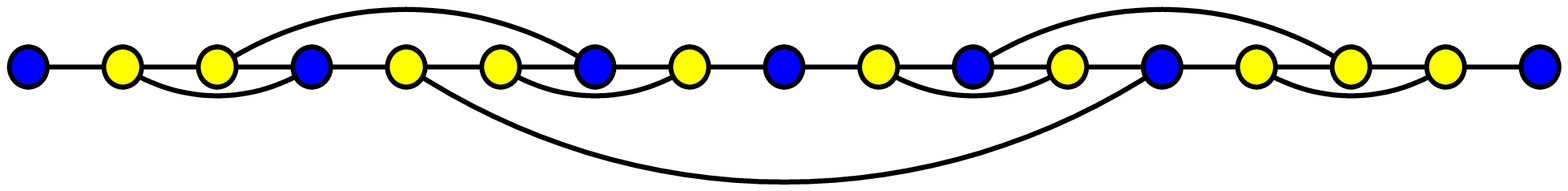}\hfill{}

\hfill{}\includegraphics[angle=-90,scale=0.4]{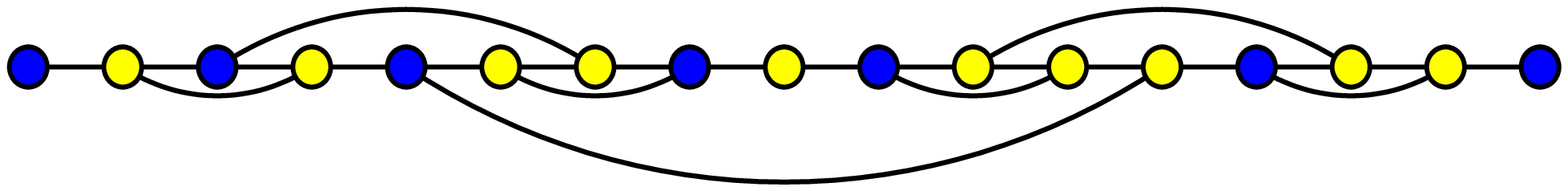}\hfill{}\includegraphics[angle=-90,scale=0.4]{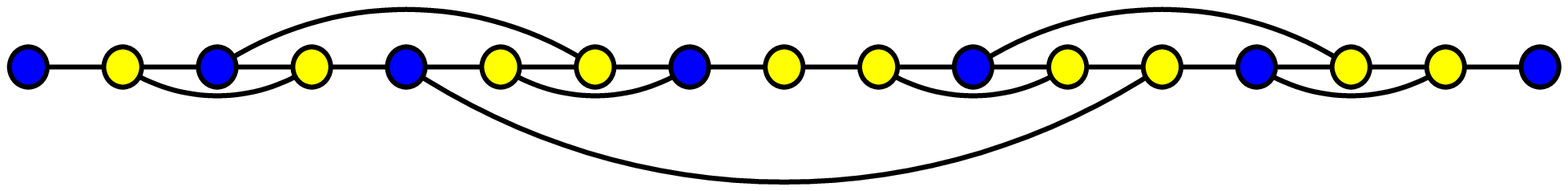}\hfill{}

\hfill{}\includegraphics[angle=-90,scale=0.4]{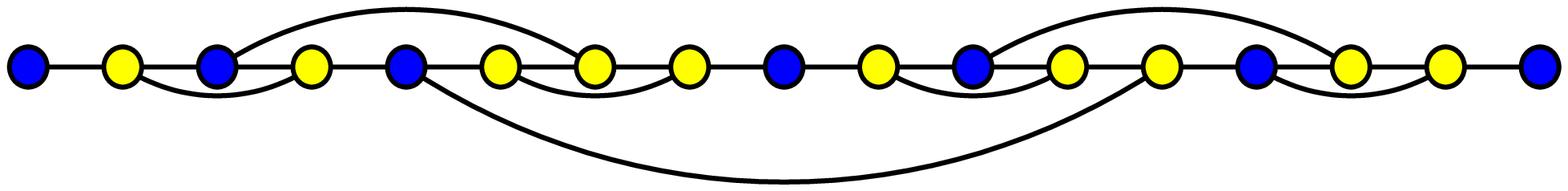}\hfill{}\includegraphics[angle=-90,scale=0.4]{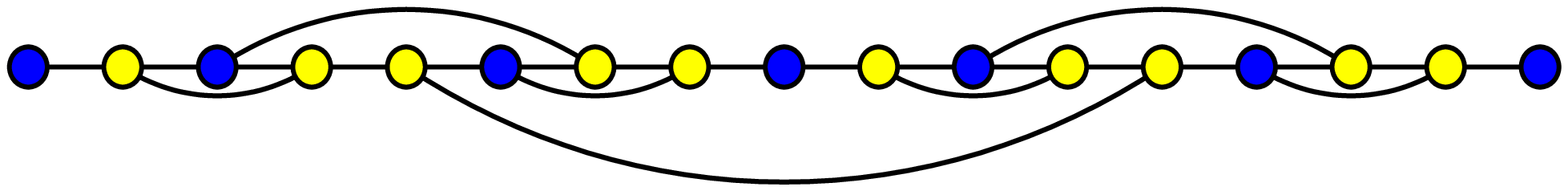}\hfill{}

\hfill{}\includegraphics[angle=-90,scale=0.4]{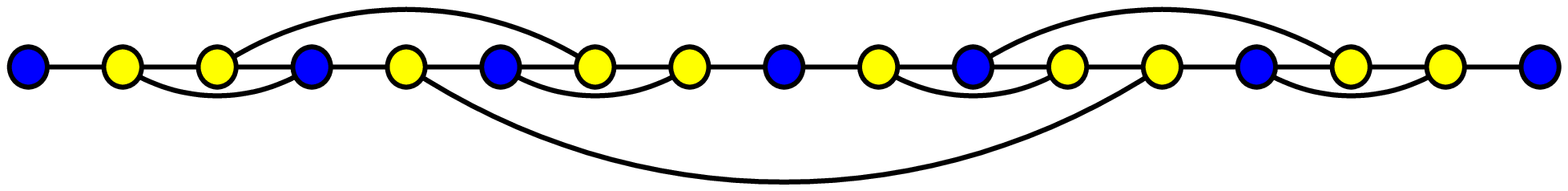}\hfill{}\includegraphics[angle=-90,scale=0.4]{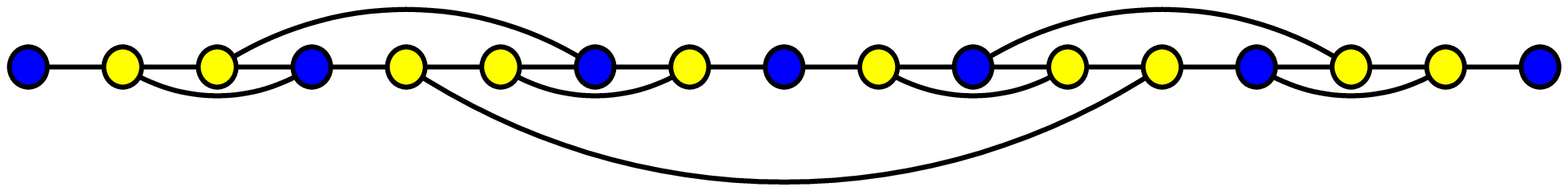}\hfill{}

\hfill{}\includegraphics[angle=-90,scale=0.4]{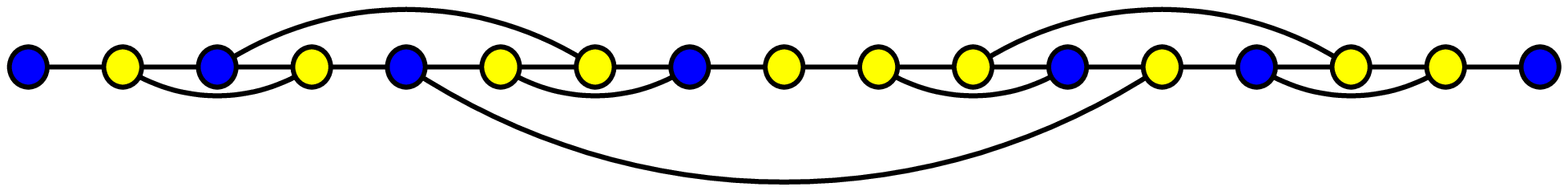}\hfill{}\includegraphics[angle=-90,scale=0.4]{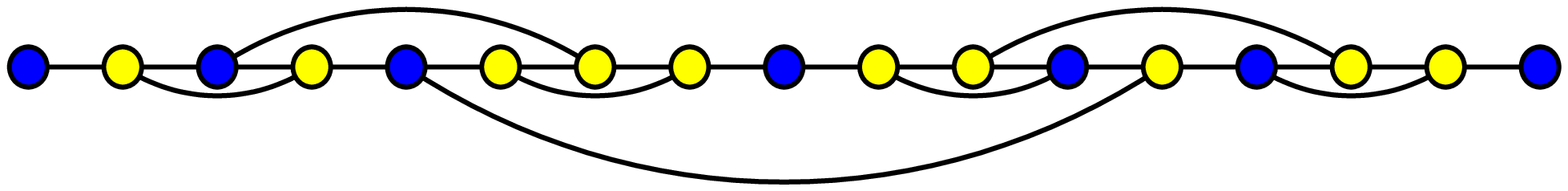}\hfill{}

\hfill{}\includegraphics[angle=-90,scale=0.4]{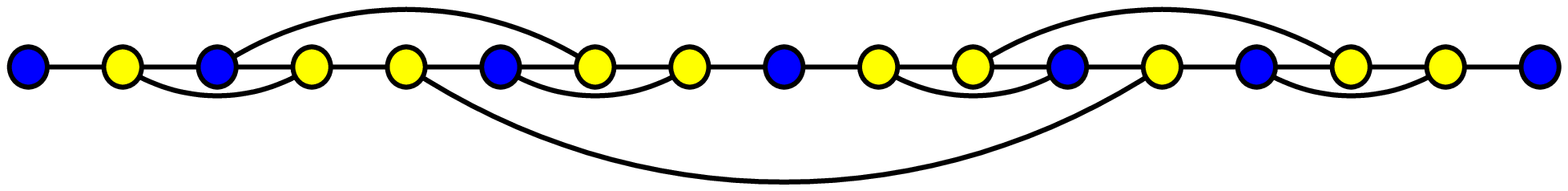}\hfill{}\includegraphics[angle=-90,scale=0.4]{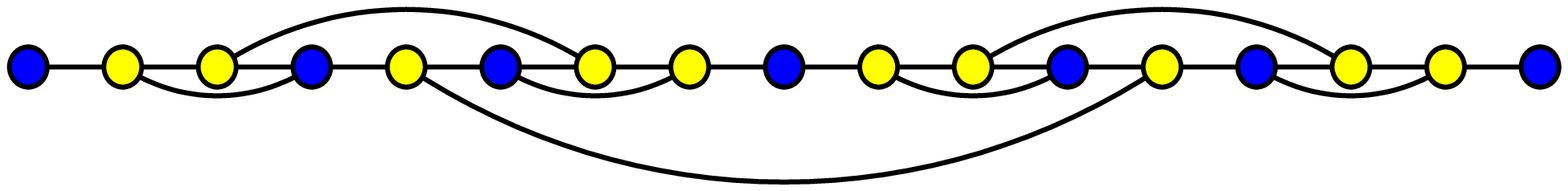}\hfill{}

\hfill{}\includegraphics[angle=-90,scale=0.4]{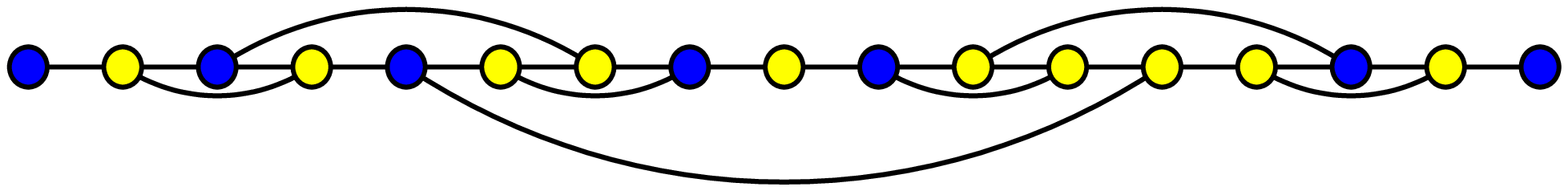}\hfill{}\includegraphics[angle=-90,scale=0.4]{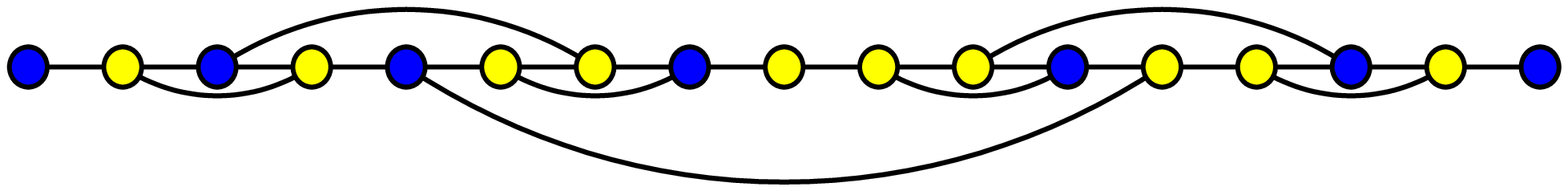}\hfill{}

\hfill{}\includegraphics[angle=-90,scale=0.4]{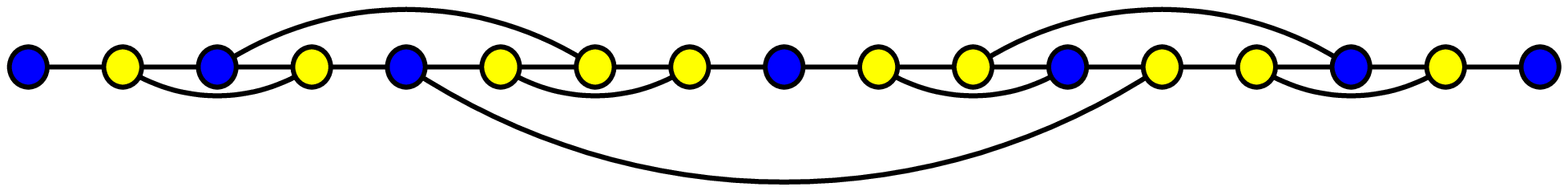}\hfill{}\includegraphics[angle=-90,scale=0.4]{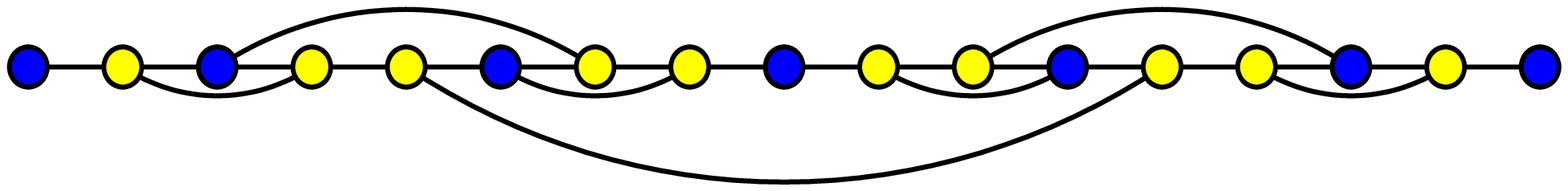}\hfill{}

\caption{\label{fig:Depiction_k4}
Depiction of perfect coverings on HN3 for $k=4$. Of all 37 solutions,
we omitted the 17 obtained by reflection from these. Light-colored
sites belong to the vertex cover, dark-colored sites mark particles
with hard-core repulsion that prevents nearest-neighbor occupation.}
\end{figure}

In the Appendix, we  derive a
partial set of recursions to approximate the number of solutions given
in Tab.~\ref{tab:sigma}. Our failure to obtain a closed set of such
equations (and an asymptotic prediction) indicates the non-trivial
origin of the entropy density. Here, we just plot the exact solutions
for $k=3$ and 4 for illustration in Figs.~\ref{fig:Depiction_k3} and
\ref{fig:Depiction_k4}. As the numerical results in
Sec.~\ref{sec:Monte-Carlo-Simulations} indicate, the optimal packing
of the lattice gas at any finite size $N=2^{k}+1$ contains for any
$k\geq3$ exactly $3\times2^{k-3}+1$ particles.


\subsubsection{Results for HN5\label{subsub:Results-for-HN5}}

\begin{figure}
\hfil\includegraphics[clip,scale=0.8]{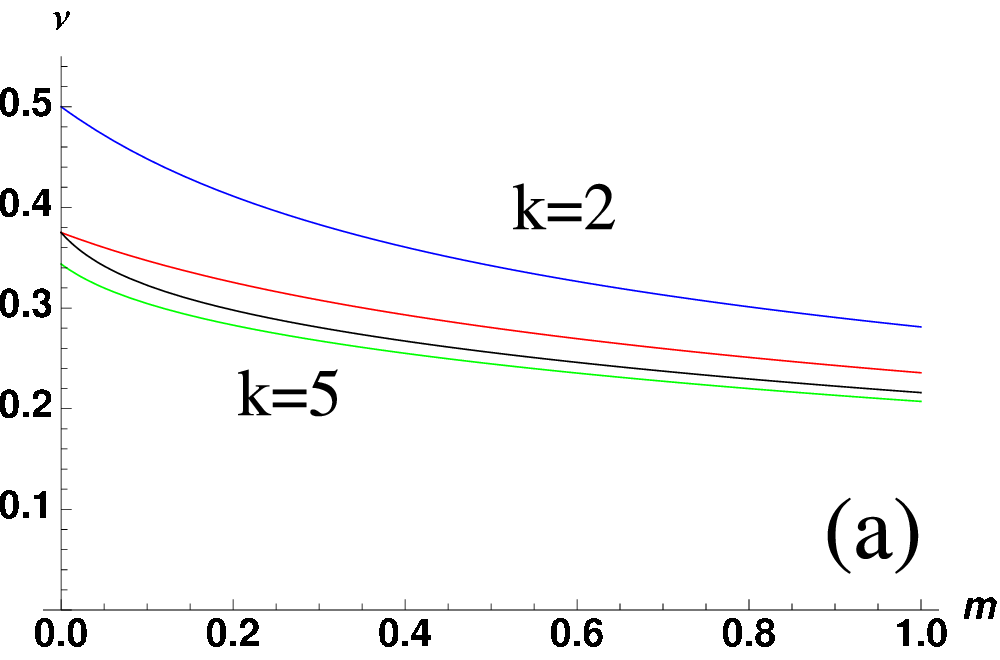}\hfil\includegraphics[clip,scale=0.8]{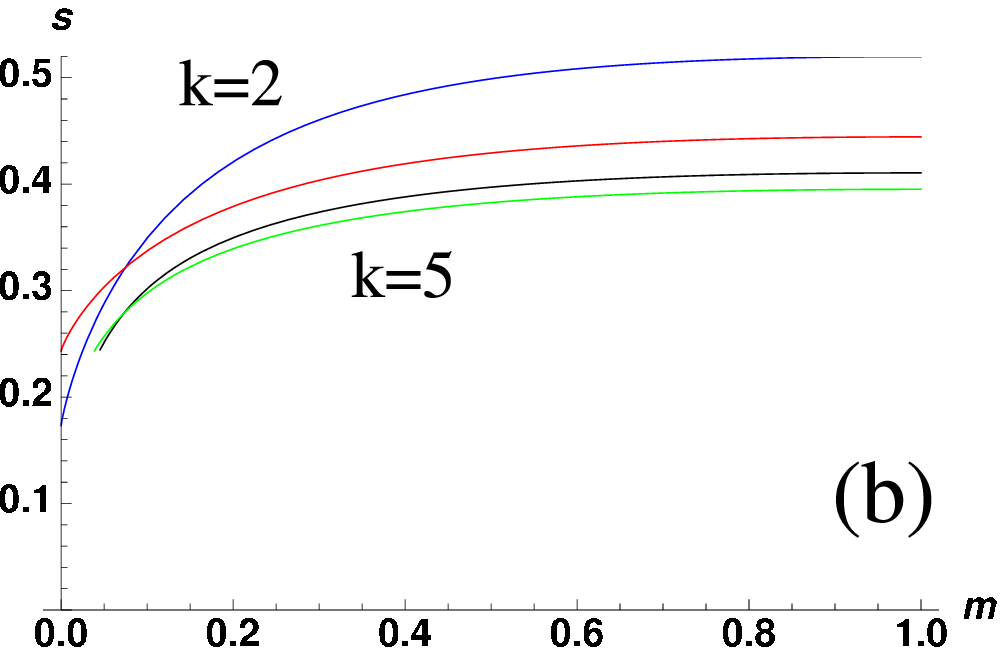}\hfil

\caption{\label{fig:entropy_HN5}
Plot of the packing fraction $\nu_{\rm VC}$ (left) and its entropy density
$s_{\rm VC}$ for the lattice-gas problem on HN5 for the first few system
sizes $N=2^{k}+1$ with $k=2,\ldots,5$ (with alternating behavior) as a
function of $m$. Each entropy drops noticeably in the limit  $m\to0$.}
\end{figure}

\begin{figure}
\includegraphics[clip]{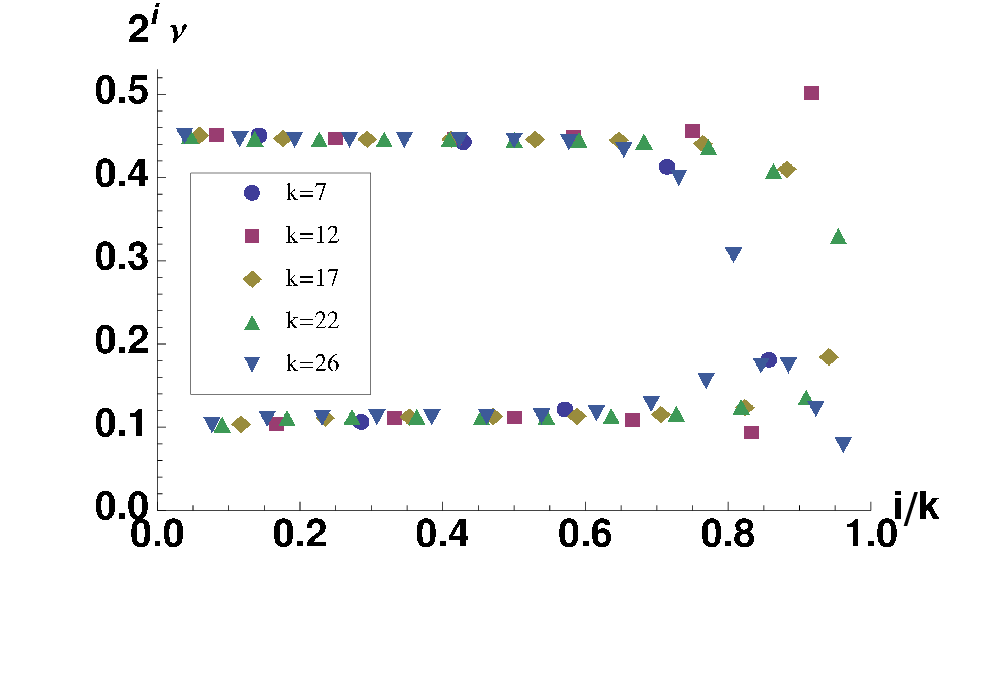}
\caption{\label{fig:levelnu_HN5}
Plot of the relative packing per level $2^{i}\nu_{i}$ on HN5 for
various system sizes $N=2^{k}+1$ with $k=7,12,17,22$, and 26, plotted
also on a relative level-scale $i/k$ at $m\to0$. In an alternating
fashion levels attain an interlaced higher or lower relative packing
(lower or higher coverage), which varies very little between the
levels and seems to converge to nontrivial values. Notice that the
apparent closing of the gap at the highest levels results from the
numerical evaluation of the RG recursions at very small but still
finite chemical activity (here, $m=10^{-9}$) . }
\end{figure}

For HN5, we find that the rudimentary partition function ${\cal
  Z}^{(1)}$ is like that for HN3 in Eq.~(\ref{eq:Z1_HN5}), except for
additional repulsive terms:
\begin{eqnarray}
{\cal Z}^{(1)} & = & C_{k-1}^{-1}\sum_{x_{0}}\sum_{x_{2^{k-1}}}\sum_{x_{2^{k}}}m_{k}^{-\left(x_{0}+x_{2^{k-1}}+x_{2^{k}}\right)}\gamma_{k-1}^{-\frac{1}{2}\left[\left(x_{0}+x_{2^{k-1}}\right)+\left(x_{2^{k-1}}+x_{2^{k}}\right)\right]}
\nonumber \\
 &  & \qquad\eta_{k-1}^{-\frac{1}{2}\left(x_{0}+x_{2^{k}}\right)}\kappa_{k-1}^{-\left(x_{0}x_{2^{k-1}}+x_{2^{k-1}}x_{2^{k}}\right)}\lambda_{k-1}^{-x_{0}x_{2^{k}}}\Delta_{k-1}^{-x_{0}x_{2^{k-1}}x_{2^{k}}}
 \label{eq:Z1_HN5}\\
 &  & \qquad\left(1-x_{0}x_{2^{k-1}}\right)\left(1-x_{2^{k-1}}x_{2^{k}}\right)\left(1-x_{0}x_{2^{k}}\right).\nonumber
 \end{eqnarray}
Hence, Eq.~(\ref{eq:lnC}) again applies, putting the focus on the
analysis of the recursion for $C_{i}$, which in its even and odd
versions read
\begin{eqnarray}
C_{2n}=\frac{\bar{\gamma}_{2n-1}}{2+\bar{\gamma}_{2n-1}}C_{2n-1}^{2}, & \quad & C_{2n-1}=\frac{m\gamma_{2\left(n-1\right)}}{2+m\gamma_{2\left(n-1\right)}}C_{2\left(n-1\right)}^{2}.
\end{eqnarray}
With the results from Sec.~\ref{subsub:Analysis-for-HN5} at hand,
when put together in the limit $m\to0$, both recursions combine into
\begin{eqnarray}
C_{2n} & \sim & mC_{2\left(n-1\right)}^{4}\left\{ A\gamma_{2n}\right\} .
\end{eqnarray}
The factor $A\gamma_{2n}$, even though it grows exponentially with $n$,
can be ignored because it does not depend on $m$. It is again easy
to sum up the logarithm of this equation (for odd values of $k$,
in this case) to get
\begin{eqnarray}
\frac{1}{2^{k}}\ln C_{k-1} & \sim & \frac{1}{8}\ln C_{2}+\frac{1}{12}\ln m\sim\frac{1}{3}\ln m,
\end{eqnarray}
with $C_{2}\sim m^{2}$ from Eqs.~(\ref{eq:IC2HN5}). As for Eq.~(\ref{eq:lnXik}),
 for the maximal packing fraction of hard-core gas particles, this implies
\begin{eqnarray}
\nu\left(\mu\to\infty\right) & = & \frac{1}{3},
\label{eq:nuHN5}
\end{eqnarray}
i.e.,  for the minimal fraction of vertices needing cover in HN5, it is
\begin{eqnarray}
c_{\rm min} & = & \frac{2}{3}.
\label{eq:mincover23}
\end{eqnarray}
In parallel to Sec.~\ref{subsub:Analysis-for-HN3}, we can  obtain only
the constant $\sigma$ defined in Eq.~(\ref{eq:Xiasymp}) for the first
few values of $k$ (see Tab.~\ref{tab:sigmaHN5}). By the same procedure
as for HN3 above, we predict here that
$s_{\rm VC}(c_{\rm min})=0.11983(1)$. For smaller system sizes we plot the
packing fraction and the entropy density for the entire range of the chemical
potential in Fig.~\ref{fig:entropy_HN5}.  Figure~\ref{fig:levelnu_HN5}
illustrates the strong alternating behavior between successive levels,
here in form of their relative packing fractions.

\begin{table}
\caption{\label{tab:sigmaHN5}
Listing of the first few values of $\sigma$ and $s_{\rm VC}$ defined in
Eqs.~(\ref{eq:Xiasymp}) and (\ref{eq:VCentropy}) for HN5 of size
$N=2^{k}+1$. The sequence for $\sigma^{N}$ soon develops non-trivial
prime factors. The entropy density for the coverage $s_{\rm VC}$
alternates and only  converges slowly to its numerically determined
limit.}
\begin{tabular}{|c|r|l|}
\hline 
$k$ & $\sigma^{N}\qquad$ & $s_{\rm VC}=\ln\sigma$\tabularnewline
\hline
2 & 2 & 0.173287\tabularnewline
3 & 7 & 0.243239\tabularnewline
4 & 6 & 0.111985\tabularnewline
5 & 159 & 0.220479\tabularnewline
6 & 1350 & 0.112623\tabularnewline
7 & 21268575 & 0.131818\tabularnewline
$\vdots$&$\vdots$&$\vdots$\tabularnewline
$\infty$ &  & 0.11983(1)\tabularnewline
\hline
\end{tabular}
\end{table}

\section{Monte Carlo Simulations\label{sec:Monte-Carlo-Simulations}}

We performed Monte Carlo simulations of the lattice gas by using the
grand canonical ensemble in Eq.~(\ref{eq:GrandPF}). To achieve a fast
convergence of the Markov chains, we used the \emph{Metropolis-Coupled
  Markov-Chain Monte Carlo} (MCMCMC) approach~\citep{geyer1991},
also termed later \emph{Parallel Tempering}~\citep{hukushima1996} in
the physics community. The idea of (MCMCMC) is to perform Monte
Carlo simulations for $n$ independent replicas studied at different
values of the chemical potential $\mu=\mu_{1},\ldots,\mu_{n}$ with
$\mu_{1}=0<\mu_{2}<\ldots<\mu_{n}$.  One allows that the replicas are
exchanged via two-replica Metropolis steps, such that an overall
detailed balance is achieved. Details of the Monte Carlo moves have been
given in previous works, e.g.~Ref.~\citep{vccluster2004}.  The
parameters for the simulations performed for this work are shown in
Tab.~\ref{tab:parameters}.

\begin{table}[ht]
 \begin{tabular}{r|rrr}
$N$  & $n$  & $\mu_{\max}$  & $t_{{\rm MCS}}$ \tabularnewline
\hline 
17  & 5  & 6  & $2\times10^{4}$ \tabularnewline
33  & 5  & 6  & $2\times10^{4}$ \tabularnewline
65  & 8  & 6  & $4\times10^{4}$ \tabularnewline
129  & 10  & 7  & $1\times10^{5}$ \tabularnewline
257  & 17  & 8  & $1\times10^{5}$ \tabularnewline
513  & 21  & 8  & $2\times10^{5}$ \tabularnewline
1025  & 33  & 10  & $1\times10^{6}$ \tabularnewline
2049  & 53  & 30  & $2\times10^{7}$ \tabularnewline
\end{tabular}
\caption{ \label{tab:parameters} 
Parameters of the MCMCMC simulations: $N$ is the system size, $n$
is the
number of different values of the chemical potential $\mu$,
$\mu_{\max}$ is the  maximum value of $\mu$, and $t_{{\rm MCS}}$ is the total number of
Monte Carlo sweeps, where in each sweep each variable is on average
allowed to flip once and $n-1$ times a replica exchange is attempted.}
\end{table}

\subsection{Monte Carlo Simulation Results}

\begin{figure}[ht]
\begin{centering}
\includegraphics[width=0.6\textwidth]{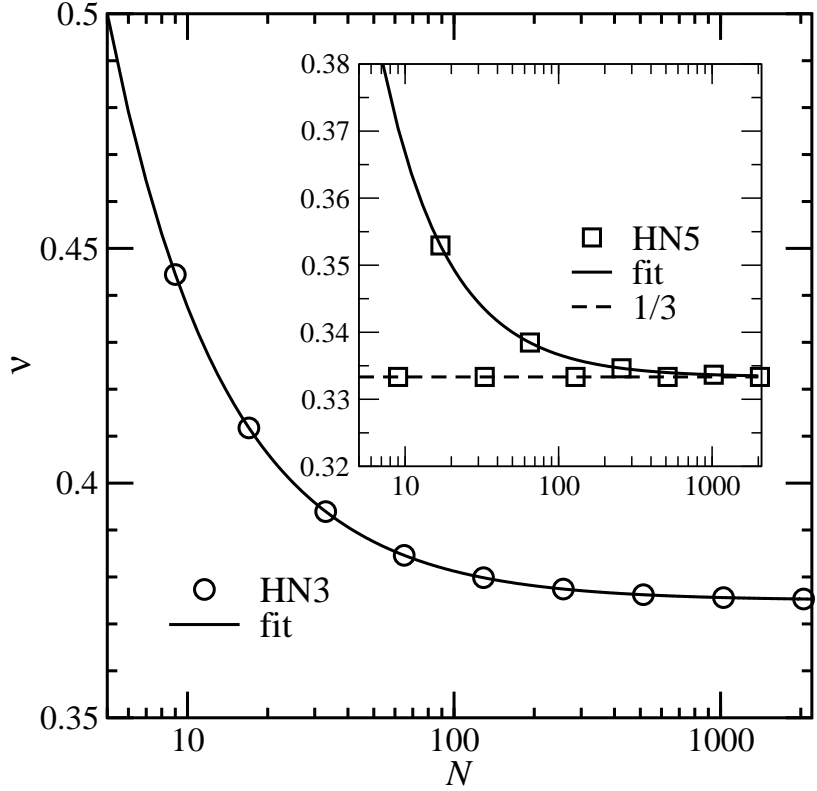} 
\par\end{centering}
\caption{ \label{fig:density_N} 
Highest density $\nu$ of the lattice gas on Hanoi networks found in
the Monte Carlo simulation as a function of system size $N$. The main
plot shows
HN3, the inset shows HN5. The solid lines represent fits to powers laws
according to Eq.~(\ref{eq:power:law}), see Tab.~\ref{tab:fits}.  The
dashed horizontal line in the inset marks the value $1/3$. }
\end{figure}

For comparison with the analytic calculations, we show the
numerical results for the density of particles. In
Fig.~\ref{fig:density_N}, the resulting largest density $\nu$,
measured at the highest value of the chemical potential $\mu$, is
shown as a function of system size $N$ for HN3 and HN5,
respectively. To extrapolate to an infinite system size, we have
fitted~\citep{practical_guide2009} the data to power laws of the form
\begin{equation}
\nu(N)=\nu_{\infty}+b\, N^{-c}\;.
\label{eq:power:law}
\end{equation}

The resulting values are displayed in Tab.~\ref{tab:fits}. Note that
for HN5, we fitted only even powers $k$, since odd powers result in
highest densities of exactly $\nu=\frac{1}{3}$. The resulting values
$\nu_{\infty}$ agree precisely with the analytical results
$\frac{3}{8}$ and $\frac{1}{3}$ for HN3 and HN5, respectively. Also the
coefficients describing the finite-size corrections seem to be
rational numbers $b=\frac{5}{8}$ and $c=-1$ for HN3 and
$b=\frac{1}{3}$ and
$c=-1$ for HN5. They can be understood in the following way, e.g., for
HN3: The number of nodes is $N=2^{k}+1$, i.e, exactly one more than a
power of 2. The number of occupied nodes for the highest density is
exactly $\frac{3}{8}$ of the $2^{k}$ nodes plus one extra node, i.e.,
$N\nu(N)=\frac{3}{8}\,2^{k}+1=\frac{3}{8}\,(2^{k}+1)+\frac{5}{8}$
which results in $\nu(N)=\frac{3}{8}+\frac{5}{8}\, N^{-1}$. In a
similar way, the scaling for the HN5 graphs can be explained, where
$N$ is not divisible by 3.

\begin{table}[ht]
 \begin{tabular}{l|lll}
 Network & $\nu_{\infty}$  & $b$  & $c$ \tabularnewline
\hline 
HN3  & 0.3750000(2)  & 0.62500(2)  & -1.00000(1) \tabularnewline
HN5 ($k$ even)  & 0.333333(7)  & 0.3333(1)  & -1.0000(1) \tabularnewline
\end{tabular}
\caption{ \label{tab:fits} 
Result of power law fits to the $\nu(N)$ data show in
Fig.~\ref{fig:density_N} according to Eq.~(\ref{eq:power:law}).  Note
that for HN5, only the data for even powers $k$ where used.}
\end{table}

Next, we go beyond the analytical calculations by studying the
properties of the solution landscape via sampling configurations of
highest density. Hence, one must ensure that configurations exhibiting
the same statistical weight in Eq.~(\ref{eq:GrandPF}) are sampled with
the same probability or frequency. For many systems exhibiting complex
solution landscapes, this is quite an
effort~\citep{ballistic2000,sg-equi2000,juanjo2003,sat_cluster2010}.

\begin{figure}[ht]
\begin{centering}
\includegraphics[width=0.6\textwidth]{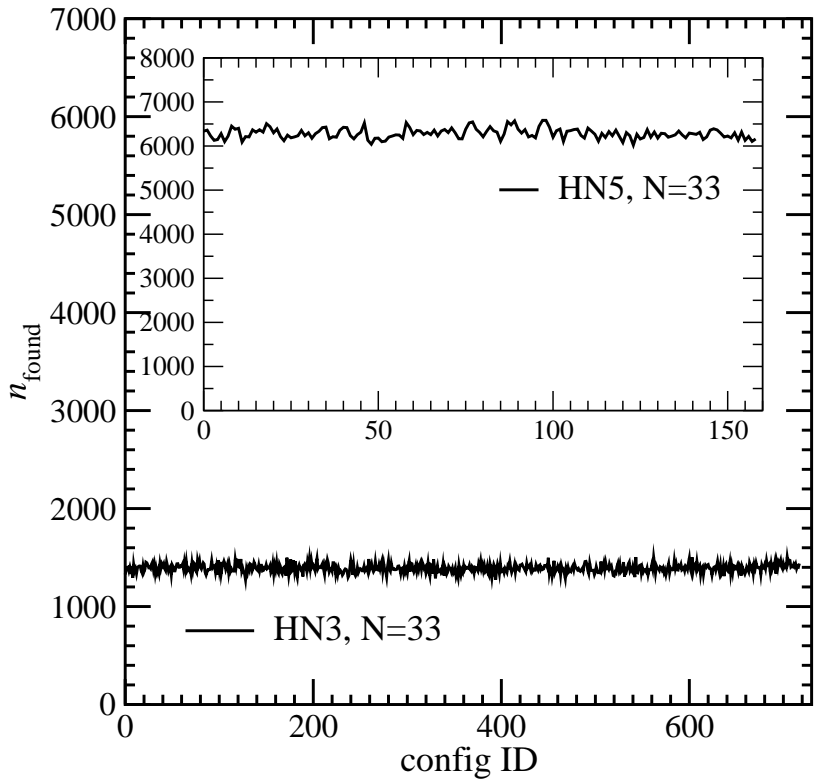} 
\par\end{centering}
\caption{ \label{fig:histo} 
Histogram of how often each configuration of highest density is sampled
during the (MC)$^{3}$ simulation of a $N=33$ node graph for HN3 (main
plot) and HN5 (inset). The total number of sampled configurations
was $10^{6}$ in both cases. }
\end{figure}

To achieve unbiased sampling here, we always stored  a configuration of
the highest density of a replica visiting the highest value $\mu_{\max}$
of the chemical potential, whenever that replica previously had
visited the value $\mu=0$ in the (MCMCMC) scheme. It may be said that the
replica has {}``performed a round trip''. This means that before a replica
is stored next time, it must again diffuse to $\mu=0$ and return to the
highest value of $\mu$~\citep{comm:neuhaus2010}. Typical round-trip
times range from around 20 for $N=17$ to around 20000 for $N=2049$.
To test whether this procedure yields unbiased sampling, we studied
small systems of size $N=33$, where, in principle, all solutions can be enumerated. For both systems, HN3 and HN5, we sampled $10^{6}$
configurations of highest density and counted how often each
configuration was found. The resulting histograms appear very flat,
see Fig.~\ref{fig:histo}.  Hence the sampling seems to work very
well, at least for Hanoi graphs.

Next, we study the configuration-landscape of the hard-core lattice
gas at the highest density. For this purpose we take, for each value
$N$ of the system size, a set of $K=200$ randomly sampled
configurations of highest density. We applied a clustering algorithm
to each set, to generate a hierarchical tree ({}``dendrogram'')
representation such that {}``similar'' configurations are grouped
closer to each other than less similar configurations. As a measure of
similarity between two configurations $\{x_{i}^{(\alpha)}\}$ and
$\{x_{i}^{(\beta)}\}$, we simply use the normalized Hamming distance
\begin{equation}
d(\{x_{i}^{(\alpha)}\},\{x_{i}^{(\beta)}\})=\frac{1}{N}\sum_{i}\delta_{x_{i}^{(\alpha)},x_{i}^{(\beta)}}\,.
\end{equation}

\begin{figure}[ht]
\begin{centering}
\includegraphics[bb=0bp 0bp 525bp 750bp,clip,width=0.7\textwidth]{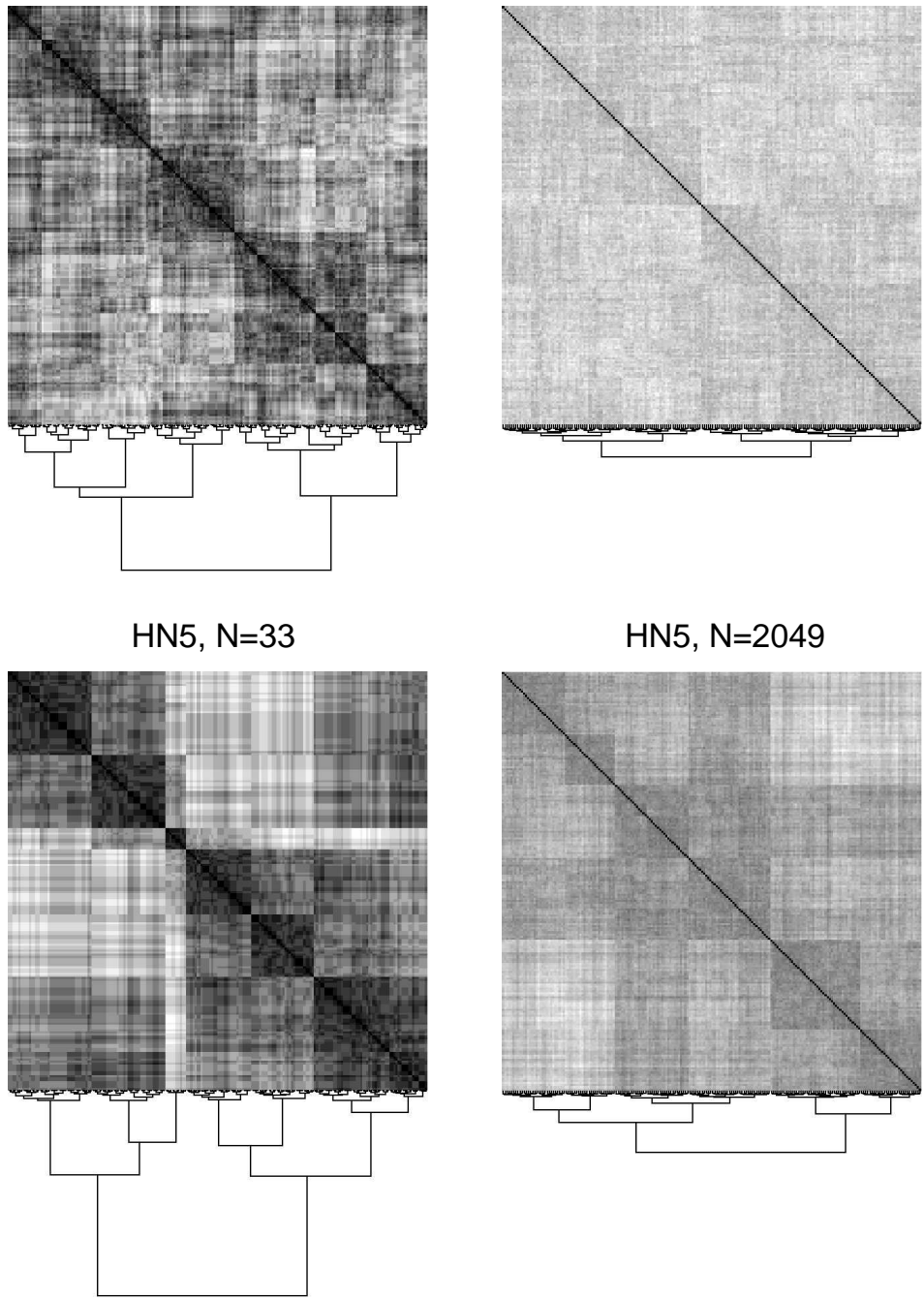} 
\par\end{centering}
\caption{ \label{fig:dendros} 
Distance-distance matrices for sets of $K=200$ randomly sampled
highest-density configurations. The columns and rows are labeled by
configurations; the order of the configurations in the rows and
columns is the same and is obtained via a clustering approach (see the
text). The clustering structure is visible by way of the trees
({}``dendrograms'') which are shown below the matrices. The entries of
each matrix are normalized Hamming distances between different
configurations, shown in gray scale (black indicates distance 0, white indicates
distance 1). }
\end{figure}

We apply the clustering algorithm of Ward~\citep{jain1988}, which
has already been applied to the analysis of phase-space structures~\citep{hed2001,vccluster2004,sat_cluster2010} (see
Ref.~\citep{hed2001} for details). The resulting dendrograms are shown
in Fig.~\ref{fig:dendros}.  The configurations are located at the
leafs of the dendrogram, at the top of each dendrogram. Arranging the
configurations from left to right as they appear in a dendrogram, a
certain order of configurations is given. Note that the order is
not unique, since for any node of the tree, the two subtrees can be
exchanged without changing the clustering.  Nevertheless, exchanging
two subtrees has no effect on the final results.  Note that any set of
vectors can be clustered and represented hierarchically in this
way. This is possible even for a set of purely random binary-valued vectors.

Whether this hierarchical clustering represents the original landscape
structure well, can be investigated in the following way. One draws
the matrix of Hamming distances by using the order of the
configurations to order the rows and columns of the matrix. If, e.g.,
one takes a set of suitably large, random binary-valued vectors, the resulting
matrices would appear basically gray, showing that the order imposed
by the clustering is artificial in this case. In
Fig.~\ref{fig:dendros} the Hamming-distance matrices are shown for a
couple of sample systems. For both cases, HN3 and HN5, at small system
sizes, a complex block-diagonal structures is visible, such that each
visible block exhibits a similar substructure.  This gives the
impression of a complex hierarchical organization of the configuration
space. Nevertheless, when going to larger system sizes, the matrices
exhibit much less contrast, which strongly indicates that for
$N\to\infty$ the solution landscape will be similar to a set of random
vectors, i.e., without any complex organization.

This result is supported when computing the \emph{cophenetic
  correlations}, which measure the correlation between the Hamming
distances $d$ and the distances $d_{c}$ along the dendrogram
\begin{equation}
{\cal K}\equiv[d\, d_{c}]-[d][d_{c}]\,,
\label{eq:cophenetic}
\end{equation}
where $[\ldots]$ is the average over pairs of configurations. Note
that $d_{c}$ is the sum of the Hamming distances along a path in the
tree connecting a pair configurations, respectively.

\begin{figure}[ht]
\begin{centering}
\includegraphics[width=0.6\textwidth]{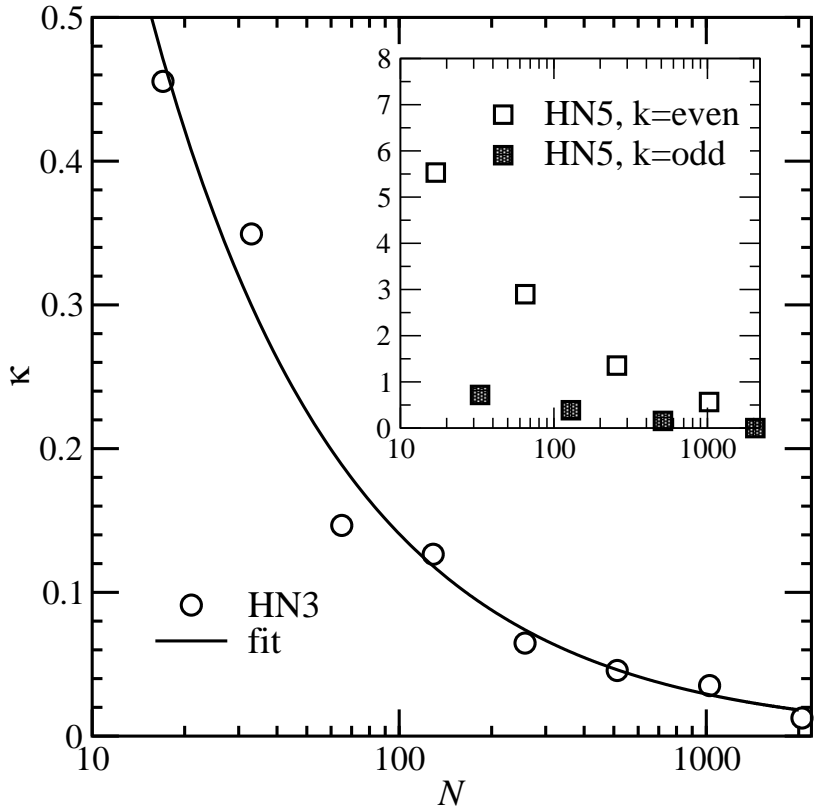} 
\par\end{centering}
\caption{ \label{fig:cophenetic} 
Cophenetic correlations in Eq.~(\ref{eq:cophenetic}) as a function
of system size for HN3 (main plot) and HN5 (inset).  The solid line
displays the function ${\cal K}(N)=3.25N^{-0.68}$. }
\end{figure}

The resulting cophenetic correlation ${\cal K}$ as a function of
system size is displayed in Fig.~\ref{fig:cophenetic}. For both cases,
HN3 and HN5, ${\cal K}$ decreases strongly as function of system size,
taking the difference between even and odd powers $k$ for HN5 into
account. For HN3, the data is compatible with a power law ${\cal
  K}(N)=3.25N^{-0.68}$. Hence, in the limit of infinite system sizes,
the hierarchical structure imposed by the clustering is not correlated
to the actual Hamming distances. This shows that the landscape of
highest-density configurations appears to be simple for both HN3 and
HN5, in strong contrast to the vertex-cover or lattice-gas problem on
random graphs~\citep{vccluster2004}.

\section{Conclusions\label{sec:Conclusions}}

We have succeeded in obtaining the optimal vertex coverage or packing
fraction for the Hanoi networks HN3 and HN5 using the renormalization
group. Our Monte Carlo simulations allowed us to confirm those results
and extend them to any finite size. We have also obtained the entropy to
arbitrary accuracy. We have shown that it is extensive and likely
non-trivial in the sense that there is no simple generator to provide
the set of all optimal configurations, a remarkable result
for such a simple, planar network.  It is even more remarkable that for each
given size the set of all possible solutions has a complex
hierarchical structure, as visible from clustering the states and
considering distance-distance matrices. Nevertheless, an analysis of the
cophenetic correlations shows that in the thermodynamic limit, a set
of random-vector-like solutions dominates entropically and makes the
solution landscape thermodynamically simple.

While there are no phase transitions in this problem, the Hanoi
networks would allow one to study analytically an interesting percolation
transition when considering an interpolation between the network's
one-dimensional backbone alone (a simple bipartite lattice with just
two perfect solutions of 1/2 coverage) and the full network (with an
extensive set of frustrated optimal solutions of coverage 5/8 for HN3
or 2/3 for HN5) by adding the small-world edges with a probability
$p$. As a  technical achievement, we derived the renormalization group
equations for hierarchy-dependent observables to obtain, for instance,
the packing fractions provided by each level of the hierarchy in the
network. Here, these observables merely reveal that higher levels of
the hierarchy become very uniform (even if alternating) in coverage,
while most of the interesting structure resides with the majority of
variables at a few lowest levels, in accordance with the numerical
study of the ultrametric relation between solutions. Nevertheless, similar
techniques might be useful to provide insights into the {}``patchy''
nature of ordering on whole classes of hierarchical networks in other
problems~\citep{Hinczewski06,Hinczewski07,Boettcher09c,Boettcher10c,Hasegawa10b}.

\section*{Acknowledgments}

SB gratefully acknowledges support from the NSF under Grant No.~DMR-0812204 and from the Fulbright Kommission for a research grant to
visit Oldenburg University, where he is deeply indebted to the
Computational Theoretical Physics Group for their kind
hospitality. AKH benefited from discussions with Thomas Neuhaus. The
simulations were performed on the GOLEM cluster of the University of
Oldenburg.

\section*{Appendix\label{sec:Appendix}}

\subsection*{A: Determining the RG-Recursion Equations\label{sub:Canonical-Partition-Function}}

In the derivation of the recursive form of the partition function in
Sec.~\ref{sec:RG-for-the}, we use Eq.~(\ref{eq:Theta}) to
transform $\Theta$ into the Ising-like form with Boolean variables $x,y,z$
\begin{eqnarray}
\Theta\left(\mu_{1},x,y,z\right) & = & 1+e^{\mu_{1}}\left(1-y\right)\left(2-x-z\right).
\nonumber \\
 & = & \exp\left\{ 2I+\frac{1}{2}G\left[\left(x+y\right)+\left(y+z\right)\right]+\frac{1}{2}H\left(x+z\right)+K\left(xy+yz\right)+Lxz+Dxyz\right\} 
\nonumber \\
& = &C_{1}^{-2}\gamma_{1}^{-\frac{1}{2}\left[\left(x+y\right)+\left(y+z\right)\right]}\eta_{1}^{-\frac{1}{2}\left(x+z\right)}\kappa_{1}^{-\left(xy+yz\right)}\lambda_{1}^{-xz}\Delta_{1}^{-xyz},
\label{eq:RG0} 
 \end{eqnarray}
where we have defined the convenient {}``activity'' parameters
\begin{equation}
C=e^{-I}, \quad \gamma=e^{-G}, \quad \eta=e^{-H}, \quad
\kappa=e^{-K}, \quad \lambda=e^{-L}, \quad \Delta=e^{-D}.
\label{eq:Activities}
\end{equation}
Equation~(\ref{eq:RG0}) matches Eq.~(\ref{eq:Theta}) for the choice of
\begin{eqnarray}
C_{1}=\frac{m_{1}}{2+m_{1}}, &  & \gamma_{1}=\frac{2+m_{1}}{m_{1}},\qquad\eta_{1}=\frac{m_{1}\left(2+m_{1}\right)}{\left(1+m_{1}\right)^{2}},
\nonumber \\
\kappa_{1}=\frac{1+m_{1}}{2+m_{1}}, &  & \lambda_{1}=\frac{\left(1+m_{1}\right)^{2}}{m_{1}\left(2+m_{1}\right)},\qquad\Delta_{1}=\frac{m_{1}\left(2+m_{1}\right)}{\left(1+m_{1}\right)^{2}},
\label{eq:RGIC}
\end{eqnarray}
(with $m_1=e^{-\mu_1}$), which serves as the initial conditions for the renormalization-group
flow for both HN3 and HN5.

\begin{figure}
\hfil\includegraphics[bb=0bp 530bp 400bp 750bp,clip,scale=0.5]{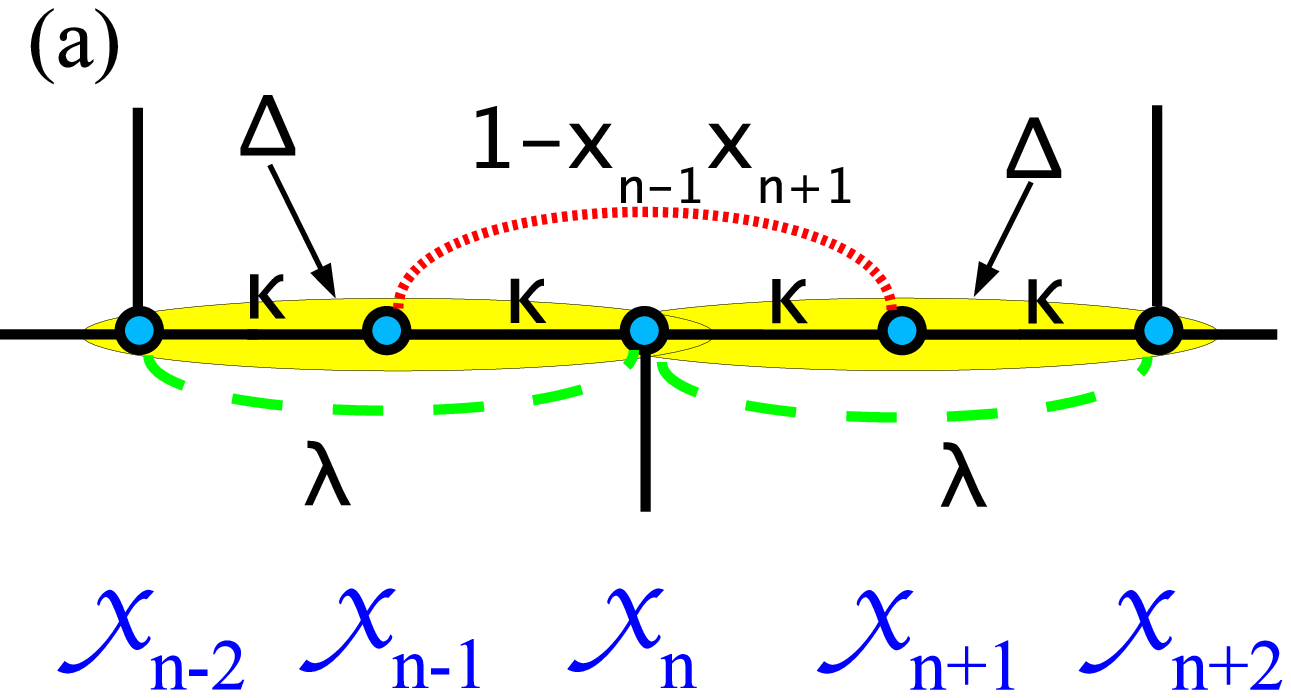}\hfil\includegraphics[bb=50bp 530bp 320bp 750bp,clip,scale=0.5]{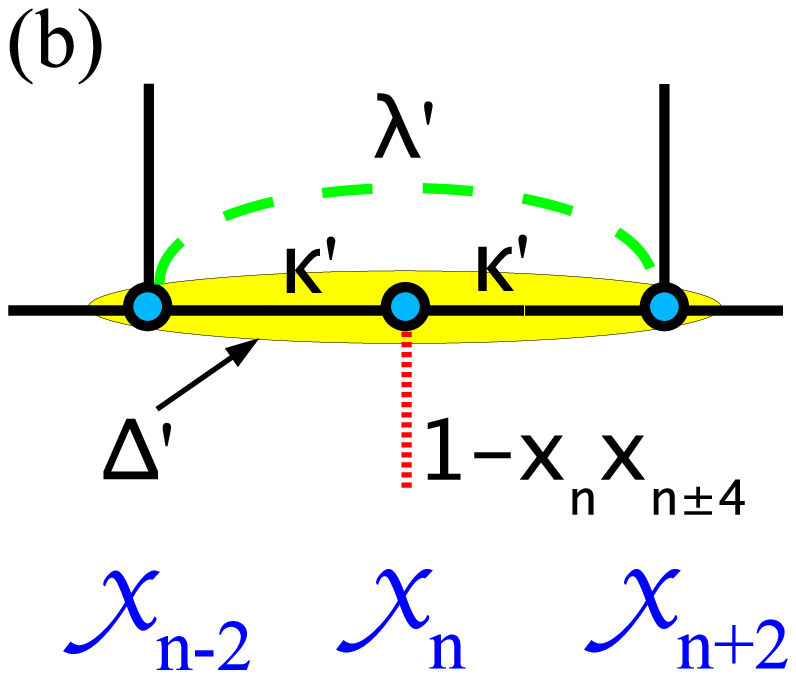}\hfil
\caption{\label{fig:RG3} 
Depiction of the graph-lets associated with the sectional partition
function $\zeta_{i}^{l}$ in Eq.~(\ref{eq:zeta}) during one RG step on
HN3. The step consists of tracing out odd-labeled variables
$x_{n\pm1}$ (taking into account the hard-core constraint relevant at
this level) in (a) and expressing the renormalized
couplings $\left(\gamma',\eta',\kappa',\lambda',\Delta'\right),$ in
(b) in terms of the old couplings
$\left(\gamma,\eta,\kappa,\lambda,\Delta\right)$.  To save space, the
one-point couplings ({}``bond magnetizations''~\citep{mckay84})
$\gamma$ and $\eta$ have been omitted. These drawings summarize the
calculations in Eqs.~(\ref{eq:zeta}) and (\ref{eq:RGrecurHN3}). }
\end{figure}

\begin{figure}
\hfil\includegraphics[bb=0bp 510bp 370bp 740bp,clip,scale=0.5]{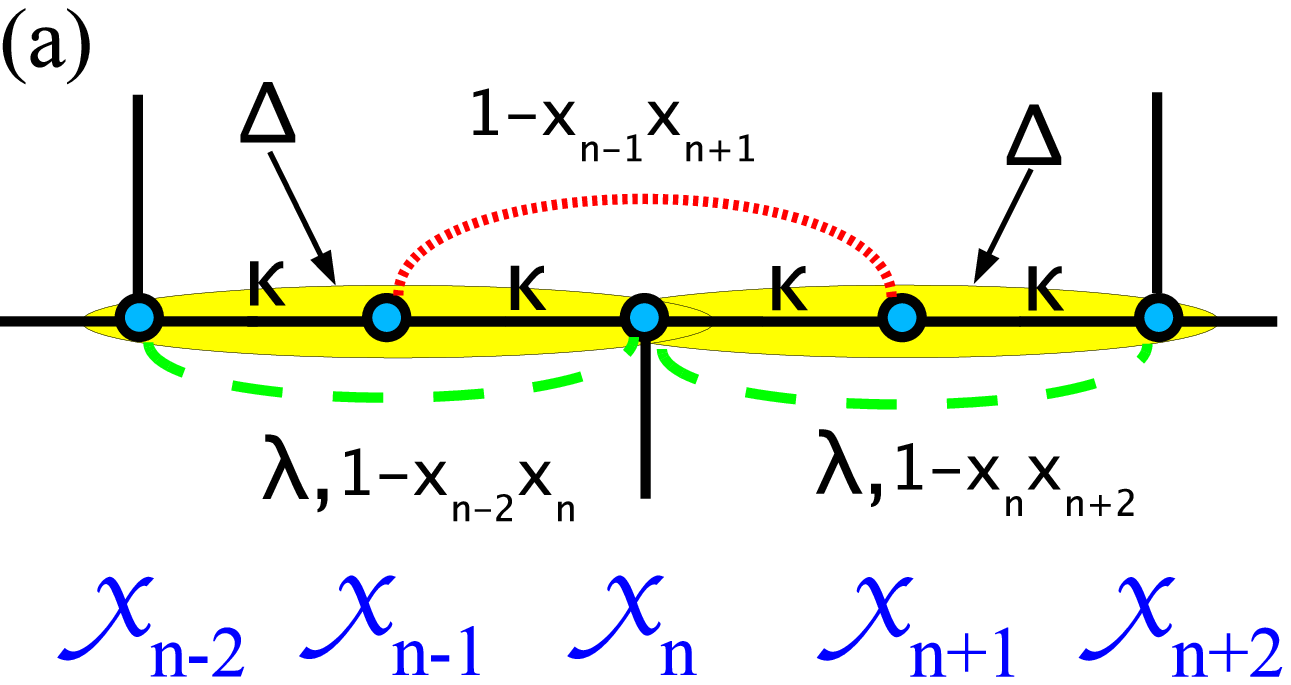}\hfil\includegraphics[bb=80bp 510bp 310bp 740bp,clip,scale=0.5]{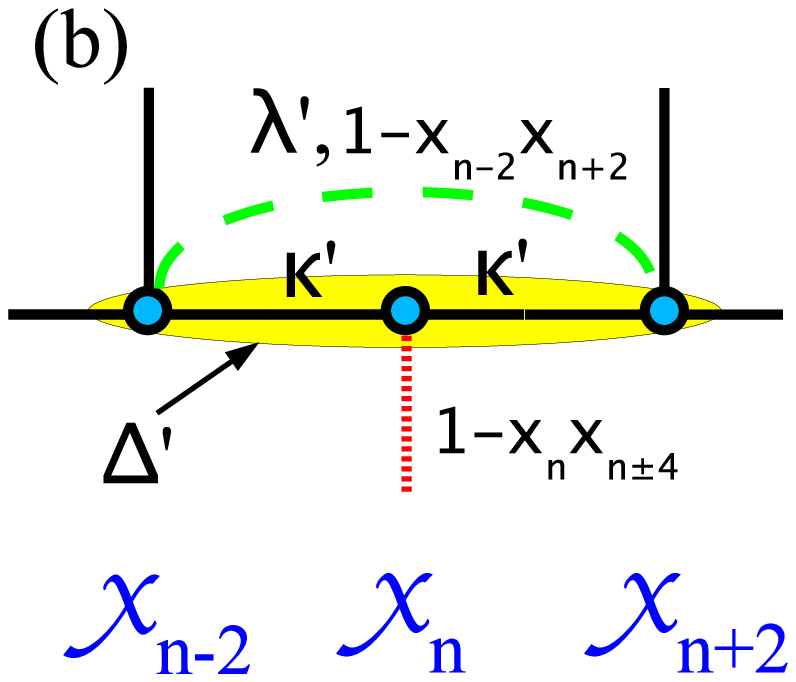}\hfil
\caption{\label{fig:RG5} 
Depiction of the (exact) RG step on HN5. This step is identical to
that for HN3 in Fig.~\ref{fig:RG3} aside from the additional hard-core
repulsive terms in (a) between $x_{n\pm2}$ and $x_{n}$  which is relevant
for the current RG step, and in (b) between $x_{n-2}$ and $x_{n+2}$ which contributes at the next level of the RG.}
\end{figure}

In terms of these renormalization-group parameters one can then show
for HN3 that the {}``sectional'' partition functions $\zeta$ have
to be written as
\begin{eqnarray}
\zeta_{i}^{l}\left(x,y,z\right) & = & \sum_{a}\sum_{b}C_{i}^{-2}m_{i+1}^{-a-b}\gamma_{i}^{-\frac{1}{2}\left[\left(x+a\right)+\left(a+y\right)+\left(y+b\right)+\left(b+z\right)\right]}\eta_{i}^{-\frac{1}{2}\left[\left(x+y\right)+\left(y+z\right)\right]}
 \nonumber \\
 &  &
 \qquad\kappa_{i}^{-\left(xa+ay+yb+bz\right)}\lambda_{i}^{-\left(xy+yz\right)}\Delta_{i}^{-\left(xay+ybz\right)}\left(1-ab\right),
\label{eq:zeta}\\
 & = & C_{i+1}^{-1}\gamma_{i+1}^{-\frac{1}{2}\left[\left(x+y\right)+\left(y+z\right)\right]}\eta_{i+1}^{-\frac{1}{2}\left(x+z\right)}\kappa_{i+1}^{-\left(xy+yz\right)}\lambda_{i+1}^{-xz}\Delta_{i+1}^{-xyz},
\nonumber
\end{eqnarray}
for which we have depicted the tracing operation graphically in Fig.~\ref{fig:RG3}.
This operation requires that, for HN3,  the renormalized quantities
at $i+1$  be expressed in terms of those at $i$ with the RG recursions
\begin{eqnarray}
C_{i+1}  = \frac{m_{i+1}\gamma_{i}C_{i}^{2}}{2+m_{i+1}\gamma_{i}}, &&
\gamma_{i+1}  =  \gamma_{i}\eta_{i}\kappa_{i}\frac{2+m_{i+1}\gamma_{i}}{2+m_{i+1}\gamma_{i}\kappa_{i}},
\nonumber \\
\eta_{i+1}  =  \kappa_{i}\frac{\left(2+m_{i+1}\gamma_{i}\right)\left(2+m_{i+1}\gamma_{i}\kappa_{i}\right)}{\left(1+\kappa_{i}+m_{i+1}\gamma_{i}\kappa_{i}\right)^{2}},
&&
\kappa_{i+1}  =  \lambda_{i}\Delta_{i}\frac{\left(2+m_{i+1}\gamma_{i}\kappa_{i}\right)\left(1+\kappa_{i}+m_{i+1}\gamma_{i}\kappa_{i}\right)}{\left(2+m_{i+1}\gamma_{i}\right)\left(1+\kappa_{i}\Delta_{i}+m_{i+1}\gamma_{i}\kappa_{i}^{2}\Delta_{i}\right)},
\label{eq:RGrecurHN3}\\
\lambda_{i+1}  =  \frac{\left(1+\kappa_{i}+m_{i+1}\gamma_{i}\kappa_{i}\right)^{2}}{\kappa_{i}\left(2+m_{i+1}\gamma_{i}\right)\left(2+m_{i+1}\gamma_{i}\kappa_{i}\right)},
&&
\Delta_{i+1}  =  \frac{\left(2+m_{i+1}\gamma_{i}\right)\left(1+\kappa_{i}\Delta_{i}+m_{i+1}\gamma_{i}\kappa_{i}^{2}\Delta_{i}\right)^{2}}{\Delta_{i}\left(2+m_{i+1}\gamma_{i}\kappa_{i}^{2}\Delta_{i}\right)\left(1+\kappa_{i}+m_{i+1}\gamma_{i}\kappa_{i}\right)^{2}}.
\nonumber 
\end{eqnarray}
For HN5, we obtain, correspondingly,
\begin{eqnarray}
\label{eq:zetaHN5}
\zeta_{i}^{l}\left(x,y,z\right) & = & \sum_{a}\sum_{b}C_{i}^{-2}m_{i+1}^{-a-b}\gamma_{i}^{-\frac{1}{2}\left[\left(x+a\right)+\left(a+y\right)+\left(y+b\right)+\left(b+z\right)\right]}\eta_{i}^{-\frac{1}{2}\left[\left(x+y\right)+\left(y+z\right)\right]}\\
 &  & \qquad\kappa_{i}^{-\left(xa+ay+yb+bz\right)}\lambda_{i}^{-\left(xy+yz\right)}\Delta_{i}^{-\left(xay+ybz\right)}\left(1-ab\right)\left(1-xa\right)\left(1-ay\right)\left(1-yb\right)\left(1-bz\right),
 \nonumber \\
& = &
C_{i+1}^{-1}\gamma_{i+1}^{-\frac{1}{2}\left[\left(x+y\right)+\left(y+z\right)\right]}\eta_{i+1}^{-\frac{1}{2}\left(x+z\right)}\kappa_{i+1}^{-\left(xy+yz\right)}\lambda_{i+1}^{-xz}\Delta_{i+1}^{-xyz}, \nonumber
\end{eqnarray}
a procedure that is graphically depicted in Fig.~\ref{fig:RG5}. Those
extra repulsion terms in HN5 then lead to dramatically simpler RG recursions
than in Eq.~(\ref{eq:RGrecurHN3}):
\begin{eqnarray}
C_{i+1}  =  \frac{m_{i+1}\gamma_{i}C_{i}^{2}}{2+m_{i+1}\gamma_{i}},&&
\gamma_{i+1}  =  \eta_{i}\frac{2+m_{i+1}\gamma_{i}}{m},\qquad
\eta_{i+1}  =  \frac{m_{i+1}\gamma_{i}\left(2+m_{i+1}\gamma_{i}\right)}{\left(1+m_{i+1}\gamma_{i}\right)^{2}},
\label{eq:RGrecurHN5}\\
\kappa_{i+1}  =  \lambda_{i}\frac{\left(1+m_{i+1}\gamma_{i}\kappa_{i}\right)}{\left(2+m_{i+1}\gamma_{i}\right)},&&
\lambda_{i+1}  = \frac{\left(1+m_{i+1}\gamma_{i}\right)^{2}}{m_{i+1}\gamma_{i}\left(2+m_{i+1}\gamma_{i}\right)},\qquad
\Delta_{i+1}  =  \frac{m_{i+1}\gamma_{i}\left(2+m_{i+1}\gamma_{i}\right)}{\left(1+m_{i+1}\gamma_{i}\right)^{2}}.
\nonumber 
\end{eqnarray}

For the discussion in Appendix~B, it
is useful to defined the vector of renormalizable parameters,
\begin{eqnarray}
\vec{A}_{i}\left(m_{1},\ldots,m_{i}\right) & = & \left(C_{i},\gamma_{i},\eta_{i},\kappa_{i},\lambda_{i},\Delta_{i}\right),\label{eq:Avector}
\end{eqnarray}
where at each level of the RG $i$ depends implicitly, through the
renormalized parameters, on the first $i$ values of the chemical potentials,
as in Eq.~(\ref{eq:RGIC2}) for the initial case $i=1$, for example.
In the analysis, we will symbolically refer to these renormalization
group equations formally as a (nonlinear) operator,
\begin{eqnarray}
\vec{A}_{i+1}\left(m_{1},\ldots,m_{i+1}\right)= & \vec{{\cal R}}_{m_{i+1}} & \left[\vec{A}_{i}\left(m_{1},\ldots,m_{i}\right)\right],
\label{eq:RGA}
\end{eqnarray}
highlighting the fact that the RG transforms depend explicitly
on the parameters $m_{i+1}$.

\subsection*{B. Hierarchical Packing Fraction\label{subsec:Hierarchical-Occ}}

For later use, we follow convention in defining  the Jacobian matrix derived
from a formal derivation of the renormalization group equations as
defined in Eqs.~(\ref{eq:Avector},\ref{eq:RGA}), 
\begin{eqnarray}
\overleftrightarrow{W}\left(\vec{A}_{i}\right) & = &
\frac{\partial\vec{A}_{i+1}}{\partial\vec{A}_{i}}=\frac{\partial\vec{{\cal
      R}}_{\mu_{i+1}}\left(\vec{A}_{i}\right)}{\partial\vec{A}_{i}}
=\frac{\partial\left(C_{i+1},\gamma_{i+1},\eta_{i+1},\kappa_{i+1},\lambda_{i+1},\Delta_{i+1}\right)}{\partial\left(C_{i},\gamma_{i},\eta_{i},\kappa_{i},\lambda_{i},\Delta_{i}\right)}.
\label{eq:DefW}
\end{eqnarray}
Using the fundamental statement for the grand partition function $\Xi^{(k)}$
of the unrenormalized system (or the free energy $f^{(k)}=2^{-k}\ln\Xi^{(k)}$,
instead) in terms of the renormalized partition functions $Z^{(i<k)}$
in Eq.~(\ref{eq:goodPFchain}), we can find for the specific packing fraction
in the $i$-th level of the hierarchy
\begin{eqnarray}
\nu_{i}\left(\vec{\mu}\right) & = & \frac{1}{2^{k}}\left\langle \sum_{j=1}^{2^{k-i}}x_{2^{i}\left(2j-1\right)}\right\rangle =\frac{\partial f^{(k)}}{\partial\mu_{i}}=-2^{-k}m_{i}\frac{d}{dm_{i}}\ln\Xi^{(k)},
\label{eq:levelnu}
\end{eqnarray}
implicitly defining the hierarchy-specific chemical potential $m_{i}=e^{\mu_{i}}$
in the form of the vector 
\begin{eqnarray}
\vec{m} & = & \left(m_{1},m_{2},\ldots,m_{k}\right).
\label{eq:mvec}
\end{eqnarray}
Applying such a derivative to the sequence in Eq.~(\ref{eq:goodPFchain}),
we obtain for $1\leq i<k$ 
\begin{eqnarray}
\frac{d}{dm_{i}}\ln\Xi^{(k)}\left(m_{1},m_{2},\ldots,m_{k}\right) & = & \frac{d}{dm_{i}}\ln Z^{(1)}\left[\vec{A}_{k-1}\left(m_{1},\ldots,m_{k-1}\right),m_{k}\right]
 =  \frac{\partial\ln Z^{(1)}\left[\vec{A}_{k-1},m_{k}\right]}{\partial\vec{A}_{k-1}}\circ\frac{d\vec{A}_{k-1}}{dm_{i}}.
 \label{eq:lnZeta}
 \end{eqnarray}
We can understand the progression of derivatives in Eq.~(\ref{eq:lnZeta})
from the result in Eq.~(\ref{eq:RGA}), 
\begin{eqnarray}
\frac{d\vec{A}_{l}}{dm_{i}} & = & \frac{d}{dm_{i}}\vec{{\cal R}}_{m_{l}}\left[\vec{A}_{l-1}\left(m_{1},\ldots,m_{l-1}\right)\right]
  =  \begin{cases}
\frac{\partial\vec{{\cal R}}_{m_{i}}}{\partial m_{i}}\left[\vec{A}_{i-1}\left(m_{1},\ldots,m_{i-1}\right)\right], & i=l,\\
\overleftrightarrow{W}\left(\vec{A}_{l-1}\right)\circ\frac{d\vec{A}_{l-1}\left(m_{1},\ldots,m_{l-1}\right)}{dm_{i}}, & i<l,\\
0, & i>l,\end{cases}
\label{eq:dAdeta}
\end{eqnarray}
using, from Eq.~(\ref{eq:DefW}), the matrix
\begin{eqnarray}
\overleftrightarrow{W}\left(\vec{A}_{l}\right) & = & \frac{\partial\vec{{\cal R}}_{m_{l+1}}}{\partial\vec{A}}\left[\vec{A}_{l}\left(m_{1},\ldots,m_{l}\right)\right].
\end{eqnarray}
Note that the distinction between the implicit and explicit derivatives
in Eq.~(\ref{eq:dAdeta}) results from the explicit occurrence of
$m_{i}$ just that once in the $i$-th RG step in the recursions
and that afterward the parameters being renormalized depend implicitly
on $m_{i}$. Thus, application of the relation in Eq.~(\ref{eq:dAdeta}),
repeatedly for all $l>i$ and once, finally, for $l=i$, yields
\begin{equation}
 \frac{d}{dm_{i}}\ln\Xi^{(k)}\left(m_{1},\ldots,m_{k}\right)
 =  \frac{\partial\ln  Z^{(1)}}{\partial\vec{A}}\left(\vec{A}_{k-1},m_{k}\right)\circ\overleftrightarrow{W}
\left(\vec{A}_{k-2}\right)\circ
\ldots\circ\overleftrightarrow{W}\left(\vec{A}_{i}\right)\circ\frac{\partial\vec{{\cal  R}}_{m}}{\partial   m}\left[\vec{A}_{i-1}\left(m_{1},\ldots,m_{i-1}\right)\right].
 \label{eq:Wchain}
\end{equation}
Now it is easy to set all chemical activities equal, $m_{i}=m$, with $1\leq i\leq k$, irrespective of which hierarchy was targeted,
to get
\begin{equation}
 \left.\frac{d}{dm_{i}}\ln\Xi^{(k)}\left(m_{1},\ldots,m_{k}\right)\right|_{m_{i}\equiv m}
  =  \begin{cases}
\frac{\partial\ln Z^{(1)}}{\partial\vec{A}}\left(\vec{A}_{k-1},m\right)\circ\overleftrightarrow{W}\left(\vec{A}_{k-2}\right)\circ\ldots\circ\overleftrightarrow{W}\left(\vec{A}_{i}\right)\circ\frac{\partial\vec{{\cal R}}_{m}}{\partial m}\left(\vec{A}_{i-1}\right), & 1\leq i<k,\\
\frac{\partial\ln Z^{(1)}}{\partial m}\left(\vec{A}_{k-1},m\right), &
i=k.\end{cases}
 \label{eq:deta}
\end{equation}

We can relate this procedure back to that for the total occupation
defined in Eq.~(\ref{eq:nu}) using a uniform $m$. To this end, we
define an extended vector of parameters with an explicit $m$-dependence
\begin{eqnarray}
\vec{A'}_{i} & = & \left(\vec{A}_{i},m\right)=\left(C_{i},\gamma_{i},\eta_{i},\kappa_{i},\lambda_{i},\Delta_{i},m\right).
\label{eq:Aprimevector}
\end{eqnarray}
Then
\begin{equation}
\frac{d}{dm}\vec{A'}_{i}  =  \left(\frac{d}{dm}\vec{A}_{i},\frac{dm}{dm}\right)
 =  \left(\overleftrightarrow{W}\left(\vec{A}_{i-1}\right)\circ\frac{d}{dm}\vec{A}_{i-1}+\frac{\partial{\cal \vec{R}}_{m}}{\partial m}\left(\vec{A}_{i-1}\right),1\right)
= \overleftrightarrow{W'}\left(\vec{A}_{i-1}\right)\circ\frac{d}{dm}\vec{A'}_{i-1},
\label{eq:DefAprime}
 \end{equation}
with the extended Jacobian matrix
\begin{eqnarray}
\overleftrightarrow{W'}\left(\vec{A}_{i-1}\right) & = & \left[\begin{array}{cc}
\frac{\partial\vec{A}_{i}}{\partial\vec{A}_{i-1}}, & \frac{\partial\vec{A}_{i}}{\partial m}\\
\frac{\partial m}{\partial\vec{A}_{i-1}}, & \frac{\partial m}{\partial m}\end{array}\right]=\left[\begin{array}{cc}
\overleftrightarrow{W}\left(\vec{A}_{i-1}\right), & \frac{\partial{\cal \vec{R}}_{m}}{\partial m}\left(\vec{A}_{i-1}\right)\\
0, & 1\end{array}\right].
\label{eq:DefWprime}
\end{eqnarray}
According to Eqs.~(\ref{eq:nu}) and (\ref{eq:levelnu}) we have $\nu=\sum_{i=1}^{k}\nu_{i}$,
so 
\begin{eqnarray}
 &  & \frac{d}{dm}\ln\Xi^{(k)}\left(m\right)
  =  \sum_{i=1}^{k}\left.\frac{d}{dm_{i}}\ln\Xi^{(k)}\left(m_{1},\ldots,m_{k}\right)\right|_{m_{i}\equiv m}\nonumber\\
 & = & \frac{\partial\ln Z^{(1)}}{\partial m}\left(\vec{A}_{k-1},m\right)+\frac{\partial\ln Z^{(1)}}{\partial\vec{A}}\left(\vec{A}_{k-1},m\right)\circ\sum_{i=1}^{k-1}\overleftrightarrow{W}\left(\vec{A}_{k-2}\right)\circ\ldots\circ\overleftrightarrow{W}\left(\vec{A}_{i}\right)\circ\frac{\partial\vec{{\cal R}}_{m}}{\partial m}\left(\vec{A}_{i-1}\right)\nonumber\\
 & = & \frac{\partial\ln Z^{(1)}}{\partial m}\left(\vec{A}_{k-1},m\right)+\frac{\partial\ln Z^{(1)}}{\partial\vec{A}}\left(\vec{A}_{k-1},m\right)\nonumber\\
 &  & \,\circ\left[\overleftrightarrow{W}\left(\vec{A}_{k-2}\right)\circ\left[\ldots\left[\overleftrightarrow{W}\left(\vec{A}_{2}\right)\circ\left[\overleftrightarrow{W}\left(\vec{A}_{1}\right)\circ\frac{\partial\vec{{\cal R}}_{m}}{\partial m}\left(\vec{A}_{0}\right)+\frac{\partial\vec{{\cal R}}_{m}}{\partial m}\left(\vec{A}_{1}\right)\right]+\frac{\partial\vec{{\cal R}}_{m}}{\partial m}\left(\vec{A}_{2}\right)\right]\ldots\right]+\frac{\partial\vec{{\cal R}}_{m}}{\partial m}\left(\vec{A}_{k-2}\right)\right]\nonumber\\
 & = & \frac{\partial\ln Z^{(1)}}{\partial m}\left(\vec{A'}_{k-1}\right)+\frac{\partial\ln Z^{(1)}}{\partial\vec{A}}\left(\vec{A'}_{k-1}\right)\circ\overleftrightarrow{W'}\left(\vec{A}_{k-2}\right)\circ\ldots\circ\overleftrightarrow{W'}\left(\vec{A}_{1}\right)\circ\frac{\partial\vec{A'}_{0}}{\partial m},
\end{eqnarray}
where the last equality follows from
Eqs.~(\ref{eq:DefAprime}) and (\ref{eq:DefWprime}).  {[}Note that
  $\frac{\partial\vec{A'}_{0}}{\partial m}=\left(0,1\right)$.{]}

\subsection*{C. Counting Optimal Packings\label{sub:Counting-Optimal-Packings}}

In this section we  attempt to determine a set of recursions to
count the number of optimal packings in HN3. In the end, we merely
succeed in providing a rigorous lower bound on the entropy density.
This exercise is interesting in its own right as it highlights the
surprising complexity in the structure of vertex covers or particle
packings on this network. The key ingredients to provide such an
approach originate with the depictions of the solutions for $k=3$ and
4 in Figs.~\ref{fig:Depiction_k3} and \ref{fig:Depiction_k4}, and with the
observation, in Sec.~\ref{sec:Monte-Carlo-Simulations}, that at each
finite system size $N=2^{k}+1$, exactly $3\times2^{k-3}+1$ particles
can be maximally packed into the network. Let us imagine we would try
to assemble the $k=4$ solutions from those of size $k=3$: We would
have to join any two solutions at one end point and add a long link
between their respective midpoints; the merging point becomes the new
midpoint and the respective open end points remain just that.  In the
process $\left(k-1\right)\to k$, we have to remove a single particle
overall, as
\begin{equation}
2\left[3\times2^{\left(k-1\right)-3}+1\right]-1=3\times2^{k-3}+1.
\label{eq:k-1tok}
\end{equation}
In this construction, it appears that only the state of midpoints and
end points is relevant, which we can denote by
$\left(n_{0}n_{\frac{N}{2}}n_{N}\right)$ with $n_{i}\in\left\{
0,1\right\} $, depending on whether that site is (1) or is not (0)
occupied by a particle. For instance, the four solutions in
Fig.~\ref{fig:Depiction_k3} would be labeled
$\left(110\right),\left(111\right),\left(101\right),\left(101\right)$
from left to right and then from top to bottom; we omit the reflection
$\left(011\right)$ of $(110)$. In fact, a glance at Fig.~\ref{fig:Depiction_k4} suggests these are the only four possibilities
realized. We have directly enumerated these classes in
Tab.~\ref{tab:HN3counts}.

\begin{table}
\caption{\label{tab:HN3counts}
Distinct classes (see the text) of solutions for HN3 for each system size
$N=2^{k}+1$. For each $k$, the total count adds up to the number of
solutions given in Tab.~\ref{tab:sigma}}
\begin{tabular}{|c|r|r|r|r|}
\hline 
$k$ & $\left(011\right)$ & $\left(110\right)$ & $\left(101\right)$ & $\left(111\right)$\tabularnewline
\hline 
3 & 1 & 1 & 3 & 2\tabularnewline
4 & 3 & 3 & 10 & 21\tabularnewline
5 & 30 & 30 & 138 & 520\tabularnewline
6 & 4140 & 4140 & 22440 & 162564\tabularnewline
\hline
\end{tabular}
\end{table}

To construct solutions of size $k$ from those at size $k-1$, we
consider all 16 pairings of these classes, which we symbolize by 
\begin{equation}
\widehat{\left(n_0n_{\frac{N}{4}}n_{\frac{N}{2}}\right)\left(n_{\frac{N}{2}}n_{\frac{3N}{2}}n_{N}\right)}_{k-1}\to\left(n_{0}n_{\frac{N}{2}}n_{N}\right)_{k},
\label{eq:solution_merge}
\end{equation}
where the over-caret corresponds to the extra long-range edge added
to connect the two former mid-points, prohibiting them from being
simultaneously occupied. With that, we find the following rules:
\begin{enumerate}
\item Merging two end-points into a new midpoint is possible 
\begin{enumerate}
\item at no cost, when both are empty, i.e.,
  $\widehat{\left(xx0)(0xx\right)}_{k-1}\to\left(x0x\right)_{k}$,
  making a new midpoint that is empty, or
\item at the expense of one particle otherwise, i.e.,
  $\widehat{\left(xx0)(1xx\right)}_{k-1}$,
  $\widehat{\left(xx1)(0xx\right)}_{k-1}$, or
  $\widehat{\left(xx1)(1xx\right)}_{k-1}\to\left(x1x\right)_{k}$.%
\footnote{One might have thought that a combination of an occupied and
  an unoccupied end point would permit the new midpoint to also be
  occupied, but it would adjoin the neighbor of
  the unoccupied end point, which is \emph{always} occupied.}
\end{enumerate}
\item Linking the two mid-points with an edge is possible
\begin{enumerate}
\item at no cost, when at least one of the two mid-points is empty, or
\item at the expense of one particle, either from the left or right mid-point,
if both mid-points are occupied.
\end{enumerate}
\end{enumerate}
The merger can proceed only when exactly one particle gets expended,
due to Eq.~(\ref{eq:k-1tok}). Hence, the combinations of 1(a) with 2(b) and 
1(b) with 2(a) are allowed. The eight permissible mergers that are
left exactly map these four classes onto themselves:
\begin{eqnarray}
\left[1.\right]\widehat{\left(011)(101\right)}_{k-1}\to\left(011\right)_{k}\qquad\left[3.\right]\widehat{\left(101)(011\right)}_{k-1}\to\left(101\right)_{k}\qquad\left[6.\right]\widehat{\left(101)(101\right)}_{k-1}\to\left(111\right)_{k}
\nonumber \\
\left[2.\right]\widehat{\left(101)(110\right)}_{k-1}\to\left(110\right)_{k}\qquad\left[4.\right]\widehat{\left(110)(101\right)}_{k-1}\to\left(101\right)_{k}\qquad\left[7.\right]\widehat{\left(101)(111\right)}_{k-1}\to\left(111\right)_{k}
\nonumber \\
\left[5.\right]\widehat{\left(110)(011\right)}_{k-1}\to\left(101\right)_{k}\qquad\left[8.\right]\widehat{\left(111)(101\right)}_{k-1}\to\left(111\right)_{k}
\label{eq:rules}
\end{eqnarray}
It seems straightforward now to deduce the recursions for the number
of configurations in each class, from one size to the next. We define
the cardinality for each set as
$x_{k}\equiv\left|\left(011\right)_{k}\right|\equiv\left|\left(110\right)_{k}\right|$,
$y_{k}\equiv\left|\left(101\right)_{k}\right|$, and
$z_{k}\equiv\left|\left(111\right)_{k}\right|$ to obtain, from the
rules in Eqs.~(\ref{eq:rules}),
\begin{eqnarray}
x_{k} & = & x_{k-1}y_{k-1},
\label{eq:count_recur}\\
y_{k} & = & 2f_{k-1}x_{k-1}y_{k-1}+2g_{k-1}x_{k-1}^{2},
\nonumber \\
z_{k} & = & y_{k-1}^{2}+2y_{k-1}z_{k-1,}
\nonumber 
\end{eqnarray}
with the initial conditions provided by Tab.~\ref{tab:HN3counts}:
$x_{3}=1,y_{3}=3,z_{3}=2$. The recursions for $x_{k}$ and $z_{k}$ are
exact, as is illustrated by evolving from one
row to the next  in Tab.~\ref{tab:HN3counts}. The recursion for $y_{k}$, though, can only provide a
lower bound on its growth. The factors of 2 in front of both terms
arises from Eq.~(\ref{eq:rules}), as the maps $\left[3.\right]$ and
$\left[4.\right]$ provide two contributions to the first while map
$\left[5.\right]$, in applying rule 2(b), gives us two ways of
removing a particle in the second term. The {}``fudge factors''
$f_{k}$ and $g_{k}$ arise because in each of these cases (and only these)
the particle removal eliminates constraints on other particles in the
respective subgraph, opening the door for an undetermined number of
further combinations from less than optimally packed subgraphs.  All
we know is that these factors are larger than unity, but they could
vary with $k$ to an unbounded size. For further analysis, we assume
that they can at least be approximated by constants $f$ and
$g$. Then, we divide the second recursion by the first  in
Eq.~(\ref{eq:count_recur}) to find $y_{k}/x_{k}\sim\lambda$ for
$k\to\infty$, with $\lambda\equiv f+\sqrt{f^{2}+2g}\geq1+\sqrt{3}$. It
is then easy to obtain asymptotically $y_{k}\sim\lambda
x_{k}\sim\left(\lambda x_{3}\right)^{2^{k-3}}$ and
$z_{k}\sim2^{k-3}\left(\lambda
x_{3}\right)^{2^{k-3}}\left(1+z_{3}/y_{3}\right)$.  The total number
of optimal packings is then $\Omega_{k}\geq2x_{k}+y_{k}+z_{k}\sim
z_{k}$, which reduces to the entropy density
\begin{equation}
s_{k}\sim\frac{\ln\Omega_{k}}{2^{k}}\geq\frac{1}{8}\ln\left(\lambda x_{3}\right)\geq\frac{\ln\left(1+\sqrt{3}\right)}{8}\approx0.1256,
\label{eq:entro_bound}
\end{equation}
using $x_{3}=1$ and the lowest value of $\lambda$. While this is a
poor lower bound, it nonetheless establishes the extensivity of the
solution-space entropy.%
\footnote{In fact, using initial conditions at $k=4,5,\ldots$ instead
  provides a monotonically increasing sequence that presumably converges to
  the exact result.} 
However, its derivation also demonstrates that the structure of optimal
packings is quite non-trivial in this network.

\bibliographystyle{apsrev}
\bibliography{/Users/stb/Boettcher}

\end{document}